# Surrogate-Assisted Inverse Design and Temperature-Dependent Electrothermal Analysis of All-Oxide Narrowband Emitter for Thermophotovoltaic Energy Conversion

Bibekananda Nath[a,b], Kawshik Nath[a,b] and Ahmed Zubair[a,*]

[a] Department of Electrical and Electronic Engineering, Bangladesh University of Engineering and Technology, Dhaka, Bangladesh
[b] Department of Electrical and Electronic Engineering, Chittagong University of Engineering and Technology, Chattogram, Bangladesh

ABSTRACT

A narrowband emitter spectrally aligned with the underlying solar cell's bandgap is essential for improving the spectral efficiency and thermal stability of high-temperature thermophotovoltaic (TPV) energy conversion systems. Emitters based on oxide materials offer a promising solution to the degradation of optical and mechanical performance caused by metal oxidation in traditional metal-dielectric emitters at elevated temperatures. Here, we presented a surrogate-assisted inverse-design framework for a narrowband 1D grating emitter comprising ITO and $Al_2O_3$ layers on a sapphire substrate. We performed Bayesian optimization over the trained ExtraTrees surrogates on a penalty-augmented objective function to acquire a high, narrowband peak emission. The surrogate model was trained on a dataset of emission spectra obtained using the finite-difference time-domain (FDTD) method. The narrowband emission's figure of merit (FOM) (($E_{\text{peak}}$), wavelength of peak emission ($\lambda_{\text{peak}}$), FWHM, in-band fraction ($f_{\text{in}}$) for determining the amount of emission outside the peak band, and concentration of peak emission near the peak) was used as the prediction target. The best optimized design exhibited $E_{\text{peak}} = 0.994$ at $\lambda_{\text{peak}} = 2181.96$ nm with FWHM $= 169.29$ nm for $t_{\text{ITO}} = 255.7$ nm, $t_{\text{Al}_2\text{O}_3} = 5.48$ nm and $P = 398$ nm which can maintain a narrowband peak emission up to approximately 50°. The inverse-designed emitter was further assessed by developing a temperature-dependent ITO dispersion model incorporating Bloch–Grüneisen-modified Drude–Lorentz parameters. Additionally, the $In_{0.73}Ga_{0.27}As$ solar cell was chosen for its bandgap (0.564 eV) alignment with the emitter's spectral emission, and achieved a maximum conversion efficiency of 25.12% for this TPV cell at an emitter temperature of 1200 K without the heat sink. Whereas incorporating active thermal management increases efficiency to approximately 32.17%. The opto-electro-thermal analysis of the proposed all-oxide emitter framework provides an efficient route for designing oxidation-resistant, spectrally selective emitters for high-temperature TPV systems capable of natural and industrial heat to electric energy conversion.

## 1. Introduction

Thermophotovoltaic (TPV) systems can directly convert heat from various natural and industrial sources into electricity. Generally, the TPV system consists of an absorber that absorbs heat from primary heat sources and is heated to a specific temperature. The back surface of the absorber is thermally connected to an emitter that emits photons to a PV cell [1–3]. However, the photon absorption of a solar cell is constrained by the bandgap of its semiconductor material [4]. Additionally, the absorption of photons with energies well above the bandgap degrades PV cell performance due to thermalization. Previous research recommended numerous solutions to address this issue. Few studies have focused on incorporating a back-surface reflector, altering cell geometry, or developing a multi-junction cell rather than a single-junction cell [5–13]. Despite their performance potential, back-surface reflectors can suffer from parasitic absorption in reflector stacks and elevated surface recombination, requiring stringent passivation, while multijunction solar cells remain limited by complex fabrication, extremely high manufacturing costs, and series-connected current-matching sensitivity to spectral variations; on the other hand, complex geometries often increase surface-area-driven recombination and impose difficult passivation and fabrication requirements, which negatively impact practical manufacturability [14–17].

To address this issue and improve spectral efficiency by tailoring the emitter's emission to the solar cell's bandgap, several studies designed a selective narrowband emitter with high optical performance and thermal stability. However, the most challenging task in achieving a narrowband response is to simultaneously optimize several competing parameters: a strong emission peak, a wavelength of peak emission to be inside the absorption range of the PV cell, a minimal FWHM with an emission range slightly above the PV cell's absorption cut-off, and very low out-of-band emission. In contrast to homogeneous materials, photonic crystal-based selective emitters are found to be a powerful route to achieving selective emission since these multilayer-periodic structures can support different types of resonance modes like Fabry–Pérot (FP) cavity and guided mode resonance and can support epsilon-near-zero effects [18–20]. Till now, various high-temperature stable refractory materials, such as molybdenum (Mo), tantalum (Ta), tungsten (W),

*Corresponding author
ahmedzubair@eee.buet.ac.bd (A. Zubair)
http://ahmedzubair.buet.ac.bd/ (A. Zubair)
ORCID(s): 0000-0001-7511-2910 (B. Nath); 0009-0001-9749-278X (K. Nath); 0000-0002-1833-2244 (A. Zubair)

platinum (Pt), silicon (Si), and iridium (Ir), based structures with various dielectric materials, have been used in photonic crystal-based emitters [21, 22]. Although refractory materials exhibit high thermal stability, they are prone to oxidation at high temperatures, which can severely degrade the structure's optical performance and thermal stability [23].

Thermal stability of the emitter can be improved by minimizing oxygen diffusion through it at elevated temperatures [24]. Using only oxide-based structures, the metal phase of the composite emitter structure can be eradicated. Recently, Song *et al* proposed all-oxide emitters based on $CeO_2$/MgO and YSZ/MgO composite multilayer structures [25]. However, their designed emitter exhibited a non-narrowband emission spectrum with low peak amplitude.

A general strategy for designing such photonic structures is forward modeling, in which the geometric parameters of a specified structure are tuned until the desired spectral response is achieved. However, such a parameter-sweep approach for high-fidelity predictions is inefficient and computationally expensive due to the large number of degrees of freedom. These challenges lead to the inverse-design method, in which the design task is framed as an optimization problem: given a specified spectral requirement, identify the structural parameters that most effectively satisfy it. Recent progress in machine learning (ML) and surrogate modeling has expedited inverse design by substituting costly repeated simulations with data-driven models that estimate the relationship between design parameters and performance metrics. After being trained on a sufficiently extensive simulation database, a surrogate model can assess potential designs significantly more quickly than full-wave simulation, enabling effective exploration of continuous parameter spaces. Bayesian optimization improves this process by probabilistically selecting new candidate designs, effectively balancing the exploitation of promising areas with the exploration of uncertain ones. By integrating physically relevant spectral metrics and constraints, surrogate-assisted Bayesian optimization provides an effective approach for designing narrowband emitters that can meet the required performance metrics.

In this work, we proposed a machine-learning-based inverse design approach to determine the optimal design of all-oxide emitter, which consists of ITO $Al_2O_3$ composite structures to achieve peak emissions exceeding 0.9. A large dataset containing the emission at discrete wavelength points ranging from 400 to 4000 nm was created by spanning the thickness of ITO (($P$, $t_{ITO}$) and $Al_2O_3$ ($t_{Al_2O_3}$) using the finite-difference time-domain (FDTD) method. Afterward, tree-ensemble surrogate models were trained to rapidly predict various design metrics. Candidate designs with a peak emissivity of ($E_{peak} \geq 0.90$) were identified by employing Bayesian optimization together with an Expected Improvement acquisition function, which provides a trade-off between bandwidth, spectral alignment, peak intensity, and sideband suppression. Further, the temperature-dependent optical response of ITO was modeled by fitting a Drude-Lorentz model to the room-temperature $n$ and $k$ spectra and introducing the temperature dependence through a Bloch–Grüneisen-based modification of the Drude damping parameter, thereby enabling evaluation of the optimized emitter under elevated-temperature operating conditions. Subsequently, the optimized emitter spectrum was coupled to an $In_{0.73}Ga_{0.27}As$ TPV cell, and its electrothermal performance was evaluated by combining optical-generation, steady-state heat-transport, and drift-diffusion simulations to determine the cell temperature, $J_{sc}$, $V_{oc}$, fill factor, maximum power density, and conversion efficiency. By unifying surrogate-assisted inverse design, temperature-resolved ITO dispersion, and cell-level electrothermal analysis, this work establishes a comprehensive computational framework for developing oxidation-resistant, spectrally selective all-oxide emitters for high-temperature TPV energy conversion.

## 2. Structure design and simulation methodology

Fig. 1 encapsulates the overall computational workflow of the TPV system used in this work. The workflow mainly consists of two coupled stages: the inverse design of the selective emitter and the electro-thermal performance evaluation of the TPV cell. First, the target optical performance metrics of the narrowband TPV emitter, such as the magnitude of $E_{peak}$ and its wavelength ($\lambda_{peak}$) within the full wavelength range, FWHM, and $f_{in}$ were defined to achieve strong narrowband emission while suppressing undesired sideband radiation. The peak emissivity and its location within the wavelength range $[\lambda_{min}, \lambda_{max}] = [400, 4000]$ nm were defined as,

$$E_{peak} = \max_{\lambda \in [\lambda_{min}, \lambda_{max}]} E(\lambda), \qquad \lambda_{peak} = \arg \max_{\lambda \in [\lambda_{min}, \lambda_{max}]} E(\lambda). \tag{1}$$

The bandwidth of peak emission was assessed by FWHM, determined from the right and left half-maximum crossings around the $\lambda_{peak}$ using linear interpolation between sampled wavelength points. Spectral selectivity was computed using a peak-centered in-band window ($\mathcal{B}$) of half-width $\Delta = 50$ nm,

$$\mathcal{B} = [\lambda_{peak} - \Delta,\ \lambda_{peak} + \Delta], \tag{2}$$

The $f_{in}$ and sideband ratio ($R_{side}$) were evaluated as follows:

$$f_{in} = \frac{\int_{\mathcal{B}} E(\lambda)\, d\lambda}{\int_{\lambda_{min}}^{\lambda_{max}} E(\lambda)\, d\lambda}, \qquad R_{side} = \frac{\int_{\lambda_{min}}^{\lambda_{max}} E(\lambda)\, d\lambda - \int_{\mathcal{B}} E(\lambda)\, d\lambda}{\int_{\mathcal{B}} E(\lambda)\, d\lambda} = \frac{1 - f_{in}}{f_{in}}. \tag{3}$$

Here, $f_{in}$ indicates how much of the total emission is concentrated near the peak. The larger the value of $f_{in}$, the more emission is within the narrowband region. On the other hand, $R_{\text{side}}$ determines the ratio of out-of-band to in-band emission, and for an efficient narrowband emitter, this parameter needs to be as low as possible. These performance metrics were used to characterize peak emission strength, spectral emission width, and out-of-band leakage, thereby enabling efficient learning and constrained optimization.

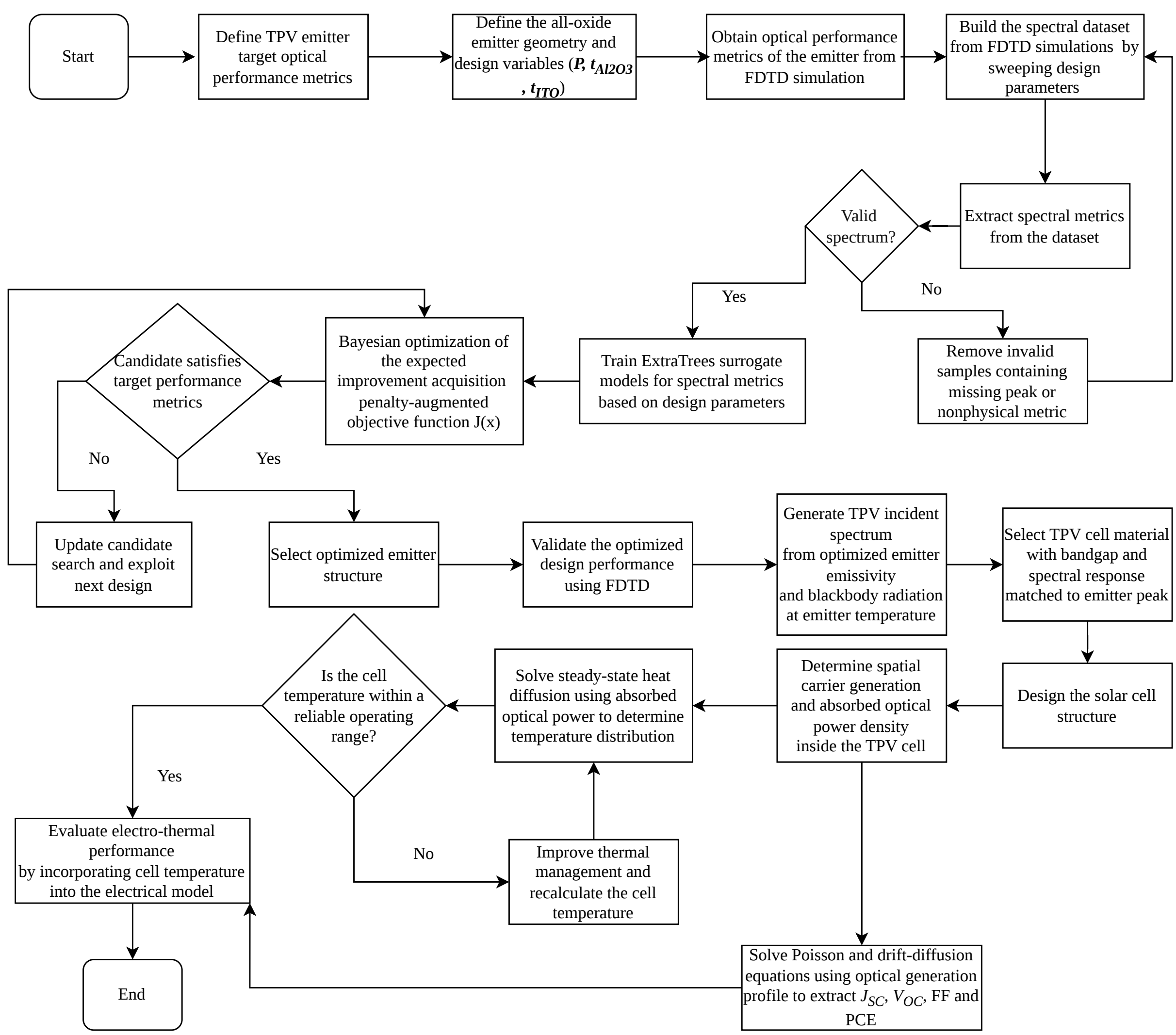

**Figure 1:** Workflow of the surrogate-assisted Bayesian optimization framework for inverse design and electro-thermal evaluation of the selective-emitter-integrated TPV system.

In this work, we developed a 1D grating structure comprised of a top $Al_2O_3$ layer and a middle ITO layer on a sapphire substrate, as can be observed from Fig. 2(a). Fig. 2(b) illustrates the FDTD simulation interface of the proposed structure in Ansys Lumerical software. The detailed FDTD source, boundary-condition, mesh, monitor, and convergence settings are summarized in Section S4 and Table S3 of the Supplementary material. The alumina top layer serves as a chemically inert and thermally stable protective coating owing to its high melting point, excellent oxidation resistance, and negligible optical loss in the near-infrared, while simultaneously facilitating the excitation of ENZ-assisted optical resonances. Alumina forms an ITO-$Al_2O_3$ solid solution at high temperatures, thereby stabilizing the ITO layer and maintaining its structural integrity. The ITO layer was selected as the active emitting medium because its free-carrier plasma response produces a strongly dispersive optical response in the infrared, including an epsilon-near-zero (ENZ) transition between dielectric-like and metal-like regimes [26]. Near the ENZ condition, the small real part of the permittivity can enhance light–matter interaction and promote strong field localization and absorption when ITO is incorporated into an appropriately designed photonic structure [27]. In contrast, wide-bandgap dielectric oxides, such as $Al_2O_3$, generally exhibit comparatively low optical loss in the near-infrared and therefore primarily provide dielectric confinement, phase control, and impedance matching rather than acting as the principal absorptive medium. The combination of a dispersive, lossy ITO layer with a low-loss $Al_2O_3$ dielectric therefore provides a suitable platform for engineering spectrally selective resonant emission. Indeed, ITO–$Al_2O_3$ architectures have previously been experimentally demonstrated for wavelength-selective thermal emission [28]. Furthermore, high-temperature studies have demonstrated that ITO thin films can retain useful optical and electrical functionality at elevated temperatures, although their properties depend on crystallinity, oxygen partial pressure, deposition conditions, and the underlying substrate [29–31]. Gregory *et al.* reported that ITO films deposited on alumina remained structurally and chemically stable even after exposure to 1400 °C (1673 K), attributing this remarkable stability to the formation of an ITO–$Al_2O_3$ interfacial region that suppresses degradation beyond the temperature predicted by bulk thermodynamics [29]. More recently, Kim *et al.* demonstrated that although the optical properties of ITO evolve with temperature because of electron–phonon interactions and oxygen-vacancy redistribution, these changes are systematic and reversible under controlled oxygen partial pressure, indicating that ITO retains its functional optical characteristics at elevated temperatures [30]. In addition, Li *et al.* reported stable electrical conductivity and microstructural stability of ITO films during prolonged operation at 1000 °C, confirming their suitability for harsh thermal environments [31]. Finally, sapphire was selected as the substrate because of its exceptional thermal stability, high melting temperature, excellent mechanical strength, low optical absorption in the near-infrared, and excellent compatibility with oxide thin-film deposition. Consequently, the $Al_2O_3$/ITO/Sapphire architecture provides an oxidation-resistant all-oxide platform in which $Al_2O_3$ ensures surface protection, ITO provides the ENZ-based selective emission mechanism, and sapphire offers a thermally robust supporting substrate. In order to determine the optimum geometrical parameters of the structures from the inverse-design method, the geometric design vector was defined as

$$\mathbf{x} = \left(P,\ t_{ITO},\ t_{Al_2O_3}\right) \tag{4}$$

here, $P$ is the structural period, $t_{ITO}$ is the ITO thickness, and $t_{Al_2O_3}$ is the $Al_2O_3$ thickness.

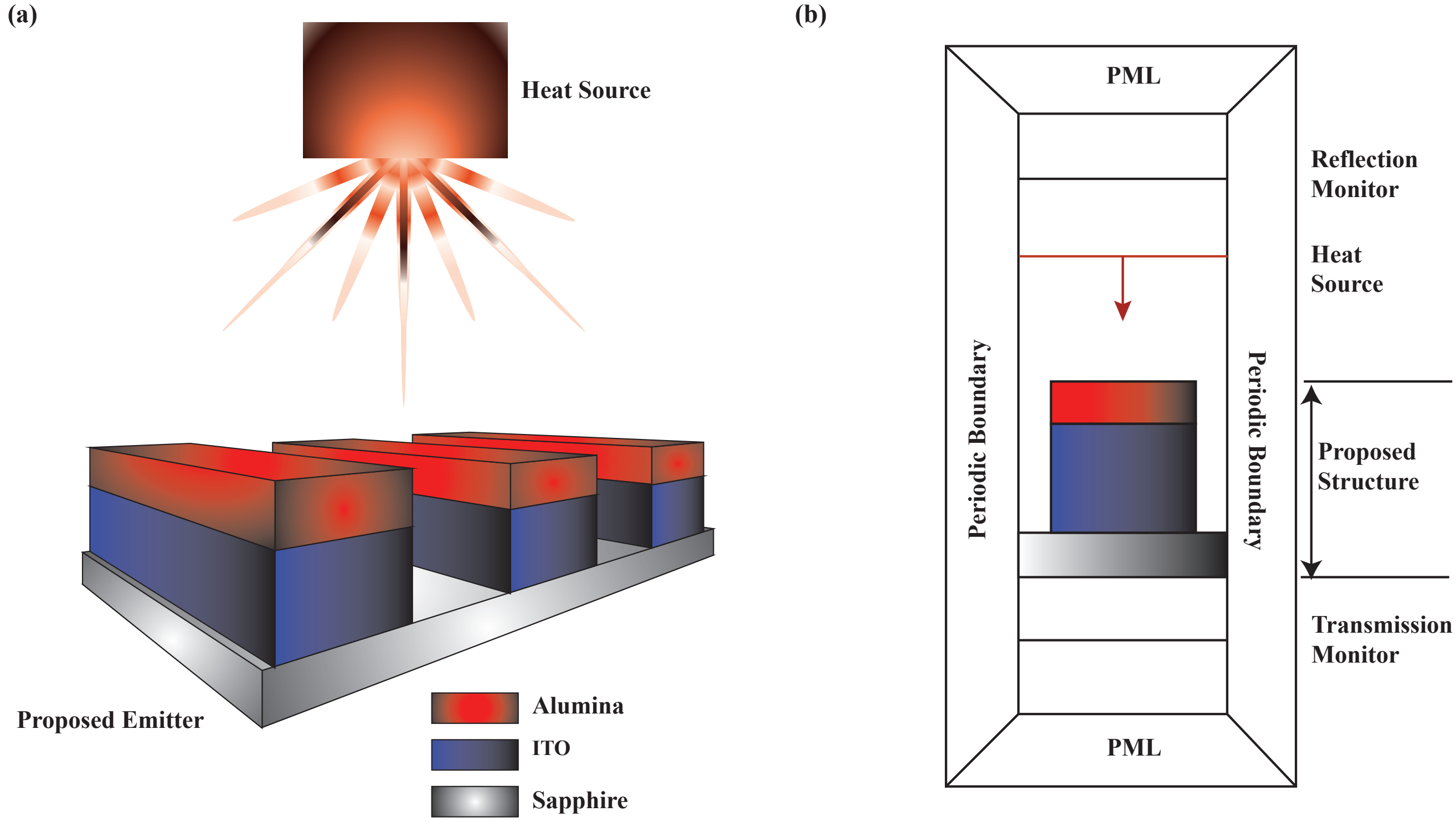

**Figure 2:** Proposed all-oxide selective emitter and numerical simulation configuration: (a) schematic of the periodic $Al_2O_3$/ITO/sapphire emitter, with structural period $P$, ITO thickness $t_{ITO}$, and $Al_2O_3$ thickness $t_{Al_2O_3}$ as the geometric design variables; and (b) FDTD computational domain showing the excitation source, reflection and transmission monitors, and a perfectly matched layer (PML) boundaries. The structure is a 1D grating with periodicity along the $x$-direction and translational invariance along the $y$-direction; periodic boundary conditions are applied in both lateral directions to represent the periodic unit cell and the effectively infinite ridge length, respectively.

After defining the design variables, a spectral dataset containing the emission over the wavelength range of 400–4000 nm was generated using FDTD simulations by varying the structural period from 340 to 460 nm, the ITO thickness from 150 to 350 nm, and the $Al_2O_3$ thickness from 5 to 350 nm. The period was sampled at nine discrete values, while the ITO and $Al_2O_3$ thicknesses were varied in 5 nm increments.

In the next step, corresponding optical performance metrics were extracted from the spectral dataset. After extracting the performance metrics, any invalid spectra with no peaks or nonphysical features were removed before model training. The detailed composition of the FDTD design library and the resulting number of valid samples are provided in Section S1 of the Supplementary Material.

In the next step, the perfectly organized dataset was used to train ExtraTrees surrogate models, which establish a mapping between structural parameters and spectral performance metrics. To predict the spectral performance of new geometries, supervised surrogate models were trained using the design vector $\mathbf{x} = (P, t_{\mathrm{ITO}}, t_{\mathrm{Al_2O_3}})$ as the input and the extracted spectral metrics as the output targets. For each target metric $m$, the surrogate model learns a mapping of the form

$$\hat{y}_m = f_m^{\mathrm{ET}}(\mathbf{x}), \tag{5}$$

where $m \in \{E_{\mathrm{peak}}, \lambda_{\mathrm{peak}}, \mathrm{FWHM}, f_{\mathrm{in}}\}$ and $f_m^{\mathrm{ET}}$ denote the ExtraTrees surrogate model trained for that metric. Here, Extremely Randomized Trees (Extra-Trees) regressors were employed due to their effectiveness in handling nonlinear data. In the ExtraTrees model, multiple randomized decision trees are trained, and the final prediction was obtained by averaging the predictions of all trees:

$$\hat{y}_m(\mathbf{x}) = \frac{1}{N_{\mathrm{tree}}} \sum_{j=1}^{N_{\mathrm{tree}}} \hat{y}_{m,j}(\mathbf{x}), \tag{6}$$

where $\hat{y}_{m,j}(\mathbf{x})$ is the prediction of the $j$-th tree for metric $m$, and $N_{\mathrm{tree}}$ is the total number of trees. Bayesian optimization was used to efficiently explore the design space, employing an expected-improvement (EI) acquisition function and a penalty-augmented objective function $J(x)$. For minimization of the scalar objective, the Expected Improvement acquisition function can be written as

$$\mathrm{EI}(\mathbf{x}) = \mathbb{E}\left[\max\left(J_{\mathrm{best}} - J(\mathbf{x}), 0\right)\right], \tag{7}$$

where $J_{\mathrm{best}}$ denotes the lowest objective-function value identified among all candidates evaluated up to the current optimization iteration. At each iteration, the next candidate is selected by

$$\mathbf{x}_{\mathrm{next}} = \arg\max_{\mathbf{x}} \mathrm{EI}(\mathbf{x}). \tag{8}$$

The optimizer generates new design candidates $\mathbf{x}$ at each iteration, the surrogate model estimates the performance metrics associated with these candidates, and a penalty-augmented scalar metric $J(\mathbf{x})$ is generated; the optimizer is then updated with evaluated $(\mathbf{x}, J)$. The loop continues until high-quality narrowband candidates are identified over the full 400–4000 nm wavelength range, with high $E_{\mathrm{peak}}$, low FWHM, and high $f_{\mathrm{in}}$. The representative objective form can be expressed as follows:

$$J(\mathbf{x}) = w_{\mathrm{F}} \widehat{\mathrm{FWHM}}(\mathbf{x}) - w_{\mathrm{E}} \hat{E}_{\mathrm{peak}}(\mathbf{x}) - w_{\mathrm{in}} \hat{f}_{\mathrm{in}}(\mathbf{x}) + \Phi(\mathbf{x}) \tag{9}$$

Here, $\Phi(\mathbf{x})$ enforces the emission peak and narrowband constraints. The peak-emissivity constraint was imposed through a penalty term,

$$\Phi(\mathbf{x}) = \begin{cases} 0, & \hat{E}_{\mathrm{peak}}(\mathbf{x}) \geq 0.90, \\ \eta\left[0.90 - \hat{E}_{\mathrm{peak}}(\mathbf{x})\right], & \hat{E}_{\mathrm{peak}}(\mathbf{x}) < 0.90, \end{cases} \tag{10}$$

where $\eta$ is a penalty coefficient. This term penalizes candidates that do not satisfy the minimum peak-emissivity requirement. Once the target criteria are satisfied, the optimized emitter geometry is selected and validated using FDTD simulations. Detailed surrogate-model hyperparameters, target transformations, sample weighting, training configuration, validation settings; the complete Bayesian-optimization implementation, including the GP-based EI acquisition procedure, objective-function weights, penalty formulation, initialization, search bounds, number of optimization evaluations, and FDTD validations, mesh-convergence analysis are provided in Section S3 and S7 of the Supplementary material.

After completing the inverse design of the all-oxide emitter, the incident spectral radiation on the TPV cell was determined by combining the optimized emitter spectral emissivity with blackbody radiation at the emitter's operating temperature. According to Planck's law, the blackbody spectral radiation at an emitter temperature is given by,

$$M_\lambda^{\mathrm{BB}}(\lambda, T_{\mathrm{emi}}) = \frac{2\pi hc^2}{\lambda^5}\left[\exp\left(\frac{hc}{\lambda k_{\mathrm{B}} T_{\mathrm{emi}}}\right) - 1\right]^{-1} \tag{11}$$

where $h$ represents the Planck's constant, $c$ denotes the speed of light, $k_{\mathrm{B}}$ is the Boltzmann constant, $\lambda$ is the wavelength, and $T_{\mathrm{emi}}$ is the emitter temperature. The incident spectral power density on the PV cell was then expressed as

$$P_{\mathrm{inc}}(\lambda, T_{\mathrm{emi}}) = F_{\mathrm{view}} \varepsilon_{\mathrm{emi}}(\lambda, \theta, \phi, T_{\mathrm{emi}}) M_\lambda^{\mathrm{BB}}(\lambda, T_{\mathrm{emi}}), \tag{12}$$

Here, $\varepsilon_{\mathrm{emi}}(\lambda, \theta, \phi, T_{\mathrm{emi}})$ indicates the spectral emissivity of the optimized emitter at a certain emitter temperature ($T_{\mathrm{emi}}$) and $F_{\mathrm{view}}$ represents the view factor between was approximated as unity to represent a closely spaced, high-view-factor configuration since the geometric view factor is governed by their relative dimensions and separation, and can approach unity when the separation is small compared with the lateral dimensions [32]. The obtained spectrum $P_{\mathrm{inc}}(\lambda, T_{\mathrm{emi}})$ was applied as the incident illumination spectrum for the optical simulation of the proposed TPV cell. In the next step, a suitable semiconductor material was selected for the solar cell based on its spectral response and bandgap, ensuring that the material's optical absorption range aligns with the emitter's peak emission. The TPV cell structure is then optically simulated to determine the spatial carrier generation rate and absorbed optical power density inside the cell.

The optical generation profile was used in the CHARGE solver of the Ansys Lumerical software, where the electrical performance parameters of the TPV cell were determined from Poisson and drift-diffusion equations as follows:

$$J_n = q\mu_n nE + qD_n \nabla n \tag{13}$$

$$J_p = q\mu_p pE - qD_p \nabla p \tag{14}$$

. Here, the electron and holes current densities measured in mA cm$^{-2}$ are denoted by $J_n$ and $J_p$, respectively. The parameter $q$ indicates the charge of an electron, while $\mu_p$ and $\mu_n$ refer to the hole and electron mobilities, respectively. $E$ represents the electric field, and $D_p$ and $D_n$ denote the hole and electron diffusion constants, respectively. The electron and hole concentrations are denoted by $n$ and $p$, respectively. These equations were solved to extract the current-voltage characteristics, short-circuit current density $J_{sc}$, open-circuit voltage $V_{oc}$, fill factor, and power conversion efficiency. In parallel, the absorbed optical power density is used as the heat source in the thermal transport model to solve the steady-state heat diffusion equation and determine the cell temperature distribution. The characteristics of these charge carriers are determined by drift and diffusion mechanisms. Additionally, this study incorporated non-radiative recombination. Therefore, the carrier dynamics were governed by both the Poisson and drift-diffusion equations, together with the electron and hole continuity equations, in which the recombination term reduces the excess carrier population and directly affects the cell's electrical performance parameters. The Shockley–Read–Hall (SRH) mechanism is the major non-radiative recombination process, which significantly limits the efficiency of TPV cell. As a result, the kinetics of carrier recombination in the absorber layer are analyzed quantitatively using the SRH recombination rate expression as follows:

$$R_{\mathrm{SRH}} = \frac{np - n_i^2}{\tau_p \left(n + n_1\right) + \tau_n \left(p + p_1\right)}, \tag{15}$$

Here, $\tau_n$ and $\tau_p$ denote the electron and hole SRH lifetimes, respectively, while $n_1$ and $p_1$ are the effective carrier concentrations associated with the trap-energy level. These quantities depend on the trap energy and semiconductor temperature.

The optical simulation provides the spatially and spectrally resolved absorbed-power density within the photoactive semiconductor region. Assuming that each absorbed photon with energy $\hbar\omega$ generates one electron-hole pair, the corresponding spectrally resolved photocarrier generation rate can be expressed as

$$g(\mathbf{r}, \omega) = \frac{p_{\mathrm{abs}}(\mathbf{r}, \omega)}{\hbar\omega}, \tag{16}$$

where $p_{\mathrm{abs}}(\mathbf{r}, \omega)$ is the local volumetric absorbed-power density at position $\mathbf{r}$ and angular frequency $\omega$, and $g(\mathbf{r}, \omega)$ denotes the corresponding spectrally resolved electron-hole-pair generation rate. The total spatial generation profile used in the carrier-transport calculation was obtained by integrating $g(\mathbf{r}, \omega)$ over the simulated spectral range.

The total spatial photocarrier-generation rate used in the carrier-transport calculation was obtained by integrating the spectrally resolved generation rate over the simulated spectrum as follows:

$$G(\mathbf{r}) = \int g(\mathbf{r}, \omega)\, d\omega \tag{17}$$

This generation rate acts as a source term in the electron and hole continuity equations. Under steady-state operation, these equations can be expressed as

$$\begin{aligned} 0 &= \frac{1}{q} \nabla \cdot \mathbf{J}_n + G(\mathbf{r}) - R(\mathbf{r}), \\ 0 &= -\frac{1}{q} \nabla \cdot \mathbf{J}_p + G(\mathbf{r}) - R(\mathbf{r}), \end{aligned} \tag{18}$$

where $R(\mathbf{r})$ represents the total carrier-recombination rate. Together with the Poisson and drift-diffusion equations, these continuity equations determine the steady-state electron and hole distributions and the resulting terminal current density. The resulting current–voltage characteristic was used to determine the short-circuit current density from

$$J_{\mathrm{sc}} = J(V = 0), \tag{19}$$

While the open-circuit voltage was obtained from the zero-terminal-current condition $J(V_{\mathrm{oc}}) = 0$. This approach accounts for the carrier transport and recombination losses and therefore provides the electrical performance of the TPV cell without assuming ideal carrier collection. The electrical power density at each operating voltage was calculated as $P(V) = |VJ(V)|$, and the maximum electrical power density, $P_{\mathrm{max}}$, was determined from the maximum of the resulting $P$–$V$ characteristic. Accordingly, the fill factor was evaluated as, $FF = P_{\mathrm{max}}/(V_{\mathrm{oc}} J_{\mathrm{sc}})$.

The wavelength-dependent absorptance of the photovoltaic structure was determined from the fraction of incident radiative power absorbed by the TPV cellas follows:

$$A_{\mathrm{cell}}(\lambda) = \frac{P_{\mathrm{abs}}(\lambda)}{P_{\mathrm{inc}}(\lambda, T_{\mathrm{emi}})}, \tag{20}$$

where $P_{\mathrm{abs}}(\lambda)$ denotes the absorbed spectral radiative power densities, respectively. The total radiative power density absorbed by the TPV cell at a given emitter temperature was then calculated as follows:

$$P_{\mathrm{abs}}(T_{\mathrm{emi}}) = \int_{\lambda_{\mathrm{min}}}^{\lambda_{\mathrm{max}}} A_{\mathrm{cell}}(\lambda) P_{\mathrm{inc}}(\lambda, T_{\mathrm{emi}})\, d\lambda, \tag{21}$$

where $\lambda_{\mathrm{min}} = 400$ nm and $\lambda_{\mathrm{max}} = 4000$ nm correspond to the simulated spectral range. Finally, the TPV cell's conversion efficiency was evaluated as follows [33]:

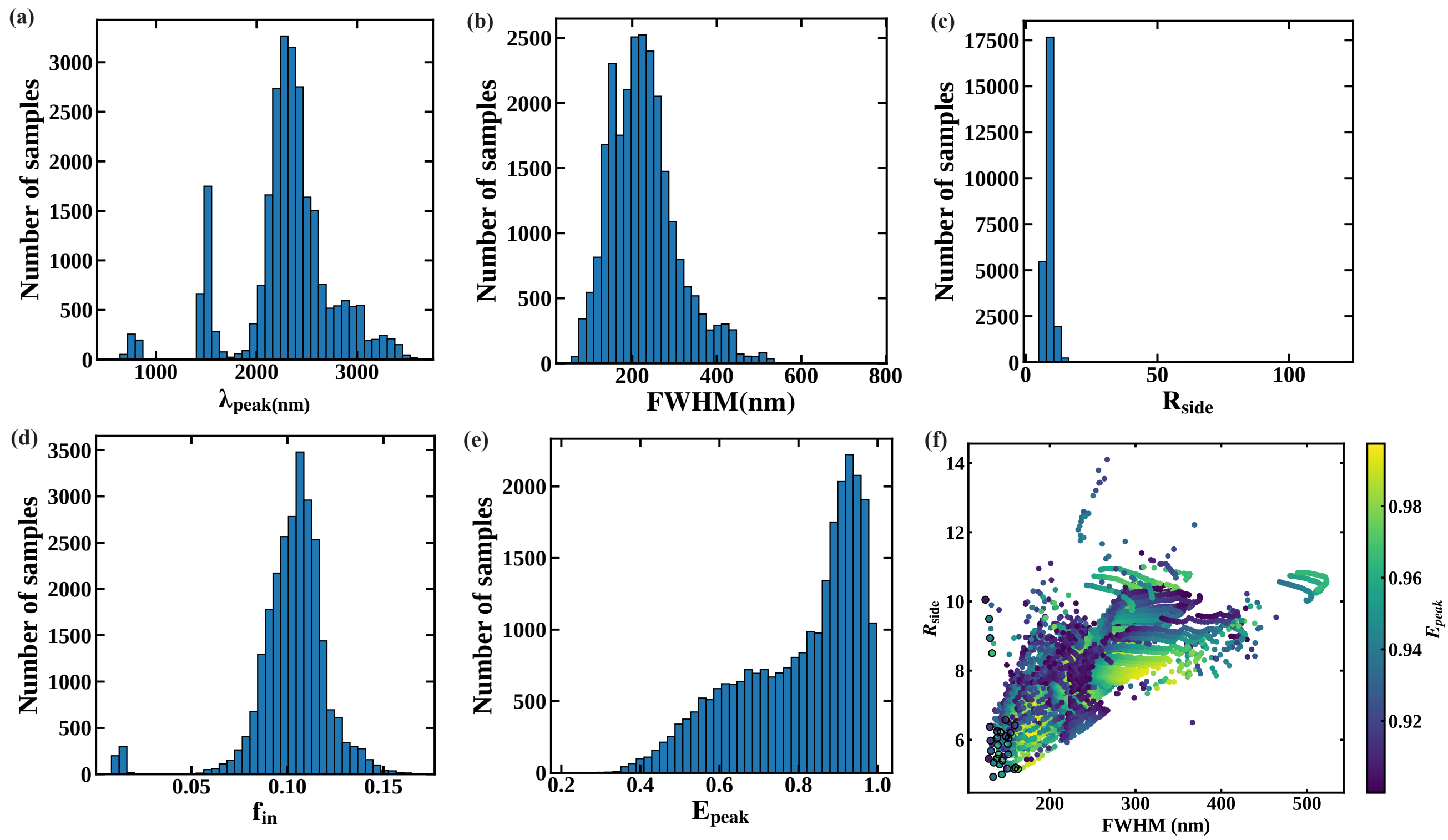

**Figure 3:** Statistical distributions of the spectrum-derived performance metrics obtained from the FDTD design library over 400–4000 nm: (a) peak wavelength $\lambda_{\text{peak}}$, (b) full width at half maximum (FWHM), (c) sideband leakage ratio $R_{\text{side}}$, and (d) in-band emission fraction $f_{\text{in}}$, (e) distribution of the $E_{\text{peak}}$ over the complete dataset and (f) FWHM versus sideband leakage ratio $R_{\text{side}}$ for candidates satisfying $E_{\text{peak}} \geq 0.90$, with marker color representing $E_{\text{peak}}$.

$$\eta_{\text{TPV}} = \frac{P_{out}}{P_{\text{abs}}(T_{\text{emi}})} = \frac{P_{out}}{P_{out} + Q_H} = \frac{FF\, V_{\text{oc}}\, J_{\text{sc}}}{P_{out} + Q_H} \quad (22)$$

Where $P_{out}$ is the total electrical output from the cell and $Q_H$ indicates the heat generated within the cell. To evaluate the temperature-dependent thermal response of the TPV cell, the spatial heat-generation profile obtained from the optical simulation was imported into the thermal transport model. The imported quantity, $Q_{\text{abs}}(\mathbf{r})$, represents the absorbed optical power converted into thermal energy within the photovoltaic structure. The steady-state temperature distribution was then obtained as follows:

$$\nabla \cdot \left(k_t \nabla T\right) + Q_{\text{abs}}(\mathbf{r}) = 0, \quad (23)$$

where $k_t$ is the thermal conductivity, and $Q_{\text{abs}}(\mathbf{r})$ is the volumetric heat-generation rate transferred from the optical simulation to the thermal model. In next step, the thermal management system was improved by modifying the heat sink material and convection condition if $T_{cell}$ exceeds the reliable operating range.

Finally, the obtained cell temperature was then incorporated into the CHARGE solver to evaluate the temperature-dependent electrical performance. In general, at elevated temperatures, the material's intrinsic carrier concentration and recombination rate increase, which enhances the dark current and mainly reduces $V_{\text{oc}}$, fill factor, maximum output power, and efficiency. The variation of the cell performance at elevated operating temperature compared to the ambient temperature was quantified as

$$\Delta X = \frac{X(T_{\text{cell}}) - X(T_{\text{amb}})}{X(T_{\text{amb}})} \times 100\%, \quad (24)$$

where $T_{\text{amb}}$ represents the ambient temperature. $X$ refers to the $J_{\text{sc}}$, $V_{\text{oc}}$, $FF$, $P_{\text{max}}$, or $\eta$.

## 3. Results and Discussion

### 3.1. Optical Performance Analysis

#### 3.1.1. *Characteristics of the FDTD-generated design space*

The period-resolved representation of the dataset obtained from the FDTD simulation is provided in Fig. S1 of the Supplementary Materials, where each panel displays the emission profile for various values ($t_{\text{ITO}}, t_{\text{Al}_2\text{O}_3}$) at a given period. From Fig. S1, the large peak emission is highly non-uniform across the design space and depends strongly on specific values of $t_{\text{Al}_2\text{O}_3}$ and $t_{\text{ITO}}$, which are determined by $P$. Except for $P = 460$ nm, the proposed structure exhibits a nearly unity peak emission for larger ITO thicknesses. Moreover, at a few periods ($P = 350, 360, 440$), the proposed structure exhibits near-unity peak emission over a broad range of $Al_2O_3$ and ITO thicknesses, whereas performance declines at lower thicknesses for other periods. This constitutes two major particulars: the first one is that the period plays an important role in facilitating or minimizing strong resonant emission, and the second one is that for a specific period, only a certain number of $t_{Al_2O_3}$, $t_{ITO}$ combinations can produce strong peak resonant emission. Besides, the diagonal-banded patterns indicate a shift in resonant emission due to interference and coupling conditions arising from layer thicknesses in multilayer periodic structures. Overall, this dataset visualization indicates that the design landscape features both structured, high-performance areas and numerous underperforming regions, highlighting the need for an efficient optimization strategy to systematically identify promising emitter geometries.

#### 3.1.2. Spectral-metric distributions and multi-objective trade-offs

Fig. 3 illustrates the distributions of the extracted performance metrics. This demonstrates the trend of pre-optimized datasets. From Fig. 3 (a), it can be observed that the peak emissivity spans a wide range of wavelengths, with a majority of the datasets having a peak emission within the 2200 nm–2500 nm wavelength range. However, the FWHM metric exhibits significant variability, as shown in Fig. 3 (b). For FWHM, numerous designs exhibit broadband peaks spanning approximately 150 nm to 300 nm. On the other hand, the sideband distribution in Fig. 3 (c) reveals that the out-of-band emission can be approximately 5 to 15 times the in-band emission. Moreover, around 5000 datasets have a sideband ratio of 5, and a maximum of around 17500 designs exhibit a sideband ratio of 10, indicating that only a small fraction simultaneously achieves a high peak with narrow width and strong sideband suppression. From Fig. 3 (d), it can be observed that most designs show an in-band fraction of around 0.11, and quite a few show a maximum in-band fraction of just over 0.16. This indicates the high concentration of emissions around the peak for a few designs of the simulated dataset. Although many datasets exhibit lower emissions, the peak emission distribution in Fig. 3 (e) indicates that around 9000 designs exhibit emissions over 90%. Together, these statistics indicate that the dataset contains promising narrowband candidates, but they occupy only a limited portion of the overall parameter space, suggesting that an efficient optimization method is required to effectively find the optimized design parameters for the required performance. Each point of Fig. 3 (f) corresponds to the datasets having various design parameters $(P, t_{ITO}, t_{Al_2O_3})$ with $E_{peak} \geq 0.90$. The x-axis represents FWHM (where smaller values indicate a narrower emission peak), and the y-axis represents the $R_{side}$. From Fig. 3, it can be said that a high peak emissivity does not uniquely determine narrowband quality: many designs achieve similar $E_{peak}$ while exhibiting substantially larger FWHM and $R_{side}$ levels. The black-outlined markers indicate the non-dominated candidates within the discrete dataset and are included only to illustrate the multi-objective trade-off between linewidth and spectral concentration. Moreover, the pronounced nonlinearity and trade-offs observed across the design space make direct selection from the sampled candidates insufficient for continuous geometric refinement. The trained surrogate models were therefore coupled with Bayesian optimization to efficiently explore the continuous design space and identify improved geometries beyond the discrete parameter combinations explicitly contained in the original dataset.

#### 3.1.3. Surrogate-model validation

For inverse design, in the first step, surrogate models were trained to predict the extracted metrics as functions of the design parameter $(P, t_{ITO}, t_{Al_2O_3})$. The predictive performance of the trained surrogate models was evaluated using three standard regression metrics: mean absolute error (MAE), root-mean-square error (RMSE), and coefficient of determination ($R^2$). For a test set containing $N$ samples, let $y_i^{FDTD}$ denote the value of a spectrum-derived metric extracted directly from the FDTD simulation, and let $y_i^{pred}$ denote the corresponding value predicted by the surrogate model. The MAE can be defined as follows:

$$\mathrm{MAE} = \frac{1}{N}\sum_{i=1}^{N}\left|y_i^{pred} - y_i^{FDTD}\right|. \tag{25}$$

The MAE gives the average absolute prediction error in the same physical unit as the predicted metric. Therefore, a smaller MAE indicates that the surrogate predictions are, on average, closer to the FDTD-derived reference values. The RMSE can be defined as follows:

$$\mathrm{RMSE} = \sqrt{\frac{1}{N}\sum_{i=1}^{N}\left(y_i^{pred} - y_i^{FDTD}\right)^2}. \tag{26}$$

Compared with MAE, RMSE places greater weight on large errors because prediction errors are squared before averaging. Thus, when RMSE is noticeably larger than MAE, it indicates the presence of a limited number of samples with relatively large prediction errors. The coefficient of determination, $R^2$, is expressed as

$$R^2 = 1 - \frac{\sum_{i=1}^{N}\left(y_i^{FDTD} - y_i^{pred}\right)^2}{\sum_{i=1}^{N}\left(y_i^{FDTD} - \overline{y}^{FDTD}\right)^2}, \tag{27}$$

where $y^{FDTD}$ is the mean value of the FDTD-derived target metric in the test set. The $R^2$ value measures the fraction of variance in the FDTD data that is captured by the surrogate model. A value close to unity indicates strong predictive agreement, whereas a lower value indicates that the surrogate does not fully reproduce the variation of that metric across the design space. Fig. 4 represents the parity plots between the FDTD-based reference values and the surrogate-predicted values for the main performance metrics. In each panel, the horizontal axis represents the metric value extracted from the FDTD spectra, while the vertical axis represents the value predicted by the surrogate model. The dashed diagonal line corresponds to ideal agreement, i.e.,

$$y^{\text{pred}} = y^{\text{FDTD}}. \tag{28}$$

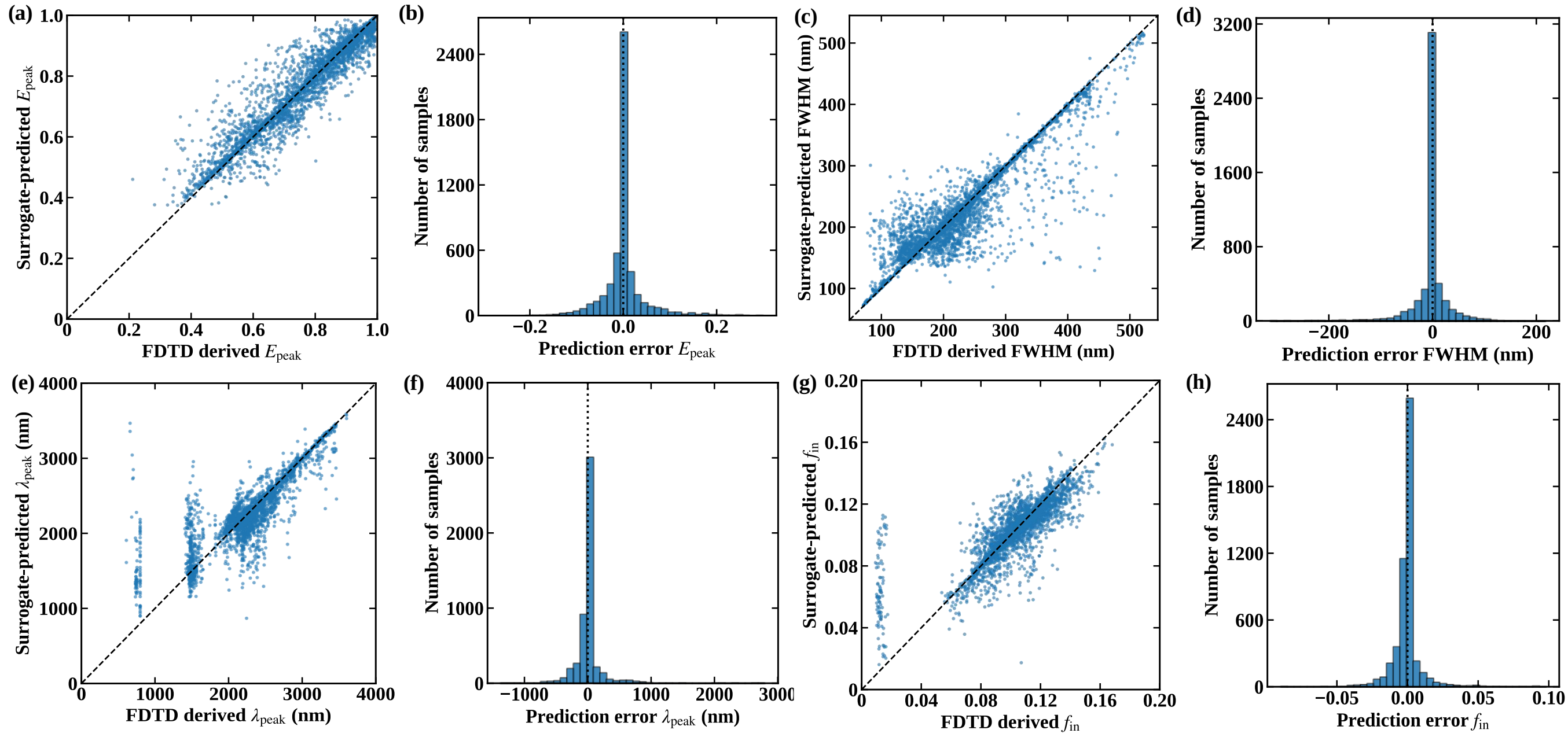

**Figure 4:** Validation of the ExtraTrees surrogate models for peak emittance and linewidth using the held-out test set: (a) parity plot of FDTD-derived and surrogate-predicted $E_{peak}$; (b) corresponding prediction-error distribution; (c) parity plot for FWHM; and (d) corresponding prediction-error distribution. The dashed diagonal lines in the parity plots represent ideal agreement, (e) parity plot of FDTD-derived and surrogate-predicted $\lambda_{peak}$; (f) corresponding prediction-error distribution; (g) parity plot for the in-band emission fraction $f_{in}$; and (h) corresponding prediction-error distribution. The dashed diagonal lines denote ideal agreement between the FDTD reference values and surrogate predictions, $y_{pred} = y_{FDTD}$.

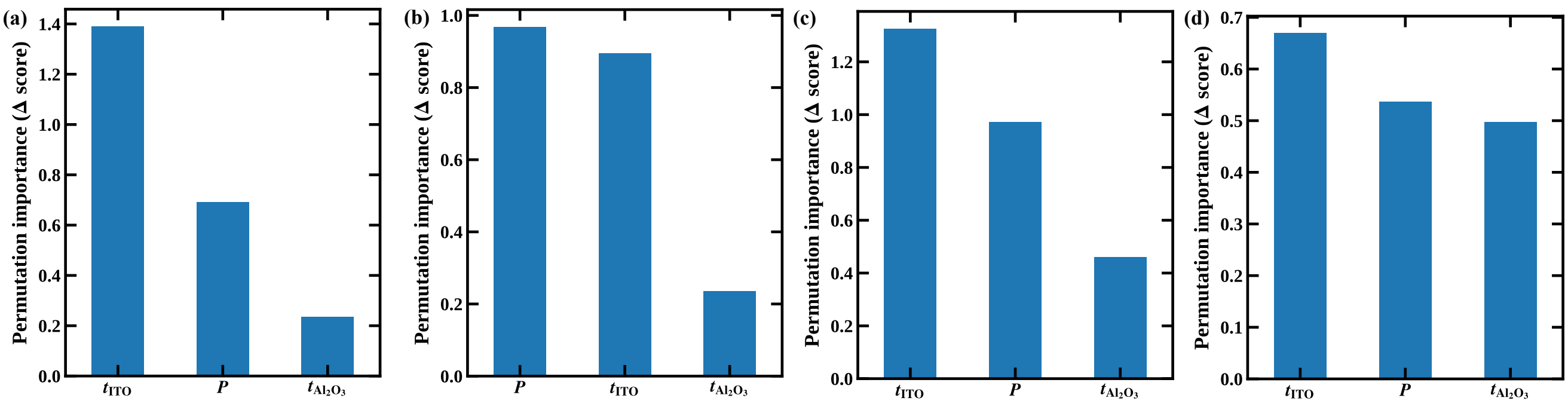

**Figure 5:** Permutation-based feature importance of the geometric parameters $P$, $t_{ITO}$, and $t_{Al_2O_3}$ for the ExtraTrees surrogate models predicting (a) $E_{peak}$, (b) $\lambda_{peak}$, (c) FWHM, and (d) $f_{in}$. The importance is quantified by the reduction in the test-set coefficient of determination, $\Delta R^2$, following random permutation of each input parameter; a larger $\Delta R^2$ indicates greater predictive dependence on the corresponding geometric variable.

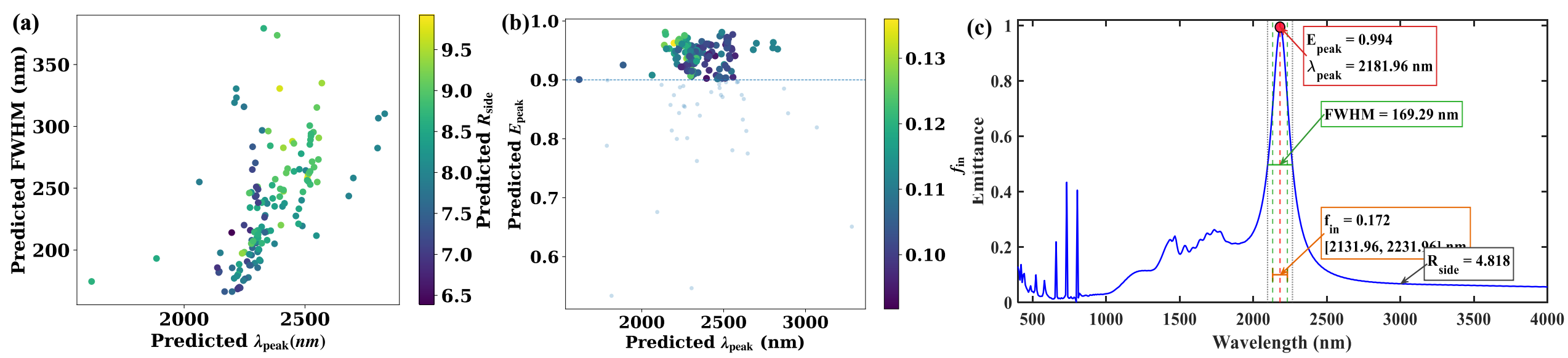

**Figure 6:** Multi-metric screening of candidate geometries generated by Bayesian optimization: (a) surrogate-predicted FWHM versus $\lambda_{peak}$, with marker color representing $R_{side}$; and (b) surrogate-predicted $E_{peak}$ versus $\lambda_{peak}$, with marker color representing the in-band fraction $f_{in}$. The horizontal dashed line in (b) denotes the imposed feasibility constraint $E_{peak} \geq 0.90$., (c)The emittance spectrum of the most optimized structure selected by surrogate-assisted Bayesian optimization.

Therefore, data points closer to the diagonal line indicate more accurate surrogate predictions.

Fig. 4(a) presents the parity plot for peak emissivity, $E_{\text{peak}}$. It can be observed that the data points are closely distributed around the ideal diagonal line over a broad emissivity range, demonstrating strong agreement between the surrogate predictions and the FDTD reference values. Quantitatively, the model achieves MAE = 0.0236, RMSE = 0.0439, $R^2$ = 0.918. The high $R^2$ value indicates that the surrogate model captures most of the variation in peak emissivity produced by changes in the geometric design variables $(P, t_{\text{ITO}}, t_{\text{Al}_2\text{O}_3})$. Since $E_{\text{peak}}$ is the primary indicator of high-emissivity resonant behavior, this result confirms that the surrogate is reliable for identifying candidate structures with strong emission peaks. Fig. 4(b) presents the prediction-error histograms for $E_{\text{peak}}$. The prediction error for each test sample is defined as

$$e_i = y_i^{\text{pred}} - y_i^{\text{FDTD}}. \tag{29}$$

A histogram centered near zero indicates that the surrogate does not show a strong systematic bias toward either overprediction or underprediction. Fig. 4(c) shows the parity plot for the full width at half maximum (FWHM). The surrogate model also predicts the linewidth with good accuracy, giving MAE = 15.6 nm, RMSE = 32.6 nm, $R^2$ = 0.837. Most data points follow the diagonal line, particularly for narrow and moderately broad resonances. This indicates that the surrogate can distinguish relatively narrowband designs from broader spectral responses. Some scattering is observed at larger linewidth values, as expected, because FWHM depends on the detailed shape of the emissivity spectrum and on the accurate determination of the half-maximum crossing points. Spectra with shoulders, multiple peaks, or broad tails can therefore lead to larger linewidth-prediction errors. Nevertheless, the relatively high $R^2$ confirms that the surrogate captures the dominant trend in linewidth across the design space. Fig. 4(d) presents the error histogram for FWHM. The error distribution is sharply centered around zero, showing that the surrogate has no strong systematic bias in linewidth prediction. The obtained values, MAE = 15.6 nm, RMSE = 32.6 nm, indicate that most linewidth predictions are close to the FDTD-derived values, while a limited number of broader or multimode spectra produce larger deviations. Since FWHM is extracted from the half-maximum points of the spectrum, it is more sensitive to spectral shape than $E_{\text{peak}}$. Nevertheless, the combined parity plot and error histogram confirm that the surrogate model can effectively distinguish narrow resonances from broader ones. Fig. 4(e) presents the parity plot for the peak wavelength, $\lambda_{\text{peak}}$. The model achieves MAE = 97.4 nm, RMSE = 218 nm, $R^2$ = 0.775. The surrogate captures the overall resonance-position trend, especially in the main wavelength region where most samples are concentrated. However, the plot also contains vertically clustered points and outliers, indicating that peak-wavelength prediction is more difficult than peak-emissivity prediction. This behavior arises because $\lambda_{\text{peak}}$ can change abruptly when the dominant resonance switches from one spectral branch to another. In such cases, two resonant modes may have comparable emissivity, and a small change in geometry can cause the extracted maximum-emissivity wavelength to jump to a different mode. Therefore, the surrogate provides a useful estimate of the resonance region, but exact peak-wavelength prediction remains more challenging for mode-switching cases. Fig. 4(f) shows the error distribution for $\lambda_{\text{peak}}$. The histogram is centered near zero, indicating that the surrogate does not exhibit a strong systematic bias in predicting the peak wavelength. However, the distribution has relatively broad tails, consistent with the larger RMSE value of 218 nm. This confirms that most peak-wavelength predictions are reasonable, but a smaller number of mode-switching samples produce large errors. These outlier cases explain why the RMSE is significantly larger than the MAE for $\lambda_{\text{peak}}$. Fig. 4(g) shows the parity plot for the in-band fraction, $f_{\text{in}}$. The model gives MAE = 0.00507, RMSE = 0.011, $R^2$ = 0.664. The points generally follow the diagonal trend, indicating that the surrogate can estimate the in-band energy fraction with moderate accuracy. However, the agreement is weaker than that obtained for $E_{\text{peak}}$ and FWHM. This is because $f_{\text{in}}$ is an integrated spectral metric and depends on both the energy contained within the selected wavelength band and the total spectral emission outside that band. Consequently, small deviations in the predicted spectral distribution can influence the extracted value of $f_{\text{in}}$. The spread observed at low true $f_{\text{in}}$ values suggests that the surrogate has more difficulty representing spectra with weak in-band confinement or poor selectivity. Fig. 4(h) shows the error histogram for $f_{\text{in}}$. The error distribution is narrow and centered near zero, with MAE = 0.00507, RMSE = 0.011. This indicates that most predictions of the in-band fraction have small absolute errors. The finite spread around zero reflects the sensitivity of $f_{\text{in}}$ to integrated spectral area, particularly when off-resonant emission contributes noticeably to the total spectral response. Overall, this result supports the use of $f_{\text{in}}$ as an auxiliary selectivity metric in the optimization process. Overall, the surrogate validation results demonstrate that the model is highly reliable for predicting $E_{\text{peak}}$, good for predicting FWHM, and moderately accurate for $\lambda_{\text{peak}}$ and $f_{\text{in}}$. Therefore, the surrogate model is used primarily for rapid design-space screening and Bayesian-optimization-guided candidate selection, while the final performance of selected structures is confirmed using direct FDTD simulations. The complete held-out validation metrics are summarized in Section 4 and Table S2 of the Supplementary Materials. Fig. 5 presents the permutation-based feature importance of the three geometric design variables ($P$, $t_{\text{ITO}}$, and $Al_2O_3$) for the surrogate models predicting $E_{\text{peak}}$, $\lambda_{\text{peak}}$, FWHM, and $f_{\text{in}}$. The analysis was performed on the test set by randomly permuting each input variable in turn while keeping the remaining variables unchanged. The importance of the $j$th parameter was quantified from the resulting reduction in the coefficient of determination as follows:

$$\Delta R_j^2 = R_{\text{base}}^2 - \left\langle R_{\text{perm},j}^2 \right\rangle, \quad (30)$$

where $R_{\text{base}}^2$ is the $R^2$ score obtained using the original unpermuted test data and $\left\langle R_{\text{perm},j}^2 \right\rangle$ is the mean score obtained after repeated random permutations of the $j$th input parameter. Accordingly, a larger $\Delta R^2$ indicates that disrupting the information contained in that parameter causes a greater deterioration in predictive performance and therefore signifies a stronger contribution to the corresponding spectral response. As shown in Fig. 5(a), $t_{\text{ITO}}$ produces the largest reduction in the predictive score for $E_{\text{peak}}$, indicating that the ITO thickness is the dominant geometric parameter governing the peak-emittance response within the investigated design space. This behavior is physically reasonable because changing the ITO thickness modifies the interaction between the resonant electromagnetic field and the optically lossy ITO layer, thereby strongly affecting the resonance strength and associated absorption. In contrast, Fig. 5(b) demonstrates the significant influence of $P$ on the prediction of $\lambda_{\text{peak}}$. The strong sensitivity of the resonance wavelength to $P$ indicates the periodic geometry's role in controlling the optical phase condition and the resonance position. Consequently, variation in the period primarily shifts the spectral location of the dominant emission mode.

The geometrical parameter importance distributions for FWHM and $f_{\text{in}}$, shown in Figs. 5 (c) and (d), further demonstrate that the spectral bandwidth and concentration of emission are governed by the combined influence of the resonant geometry and the ITO layer. Changes in these parameters modify both the spectral position and damping of the supported resonance, thereby affecting the linewidth and the fraction of emission concentrated around the principal peak. In comparison, $t_{\text{Al}_2\text{O}_3}$ exhibits a relatively smaller contribution to most of the predicted responses, suggesting that, within the investigated thickness range, it primarily provides a secondary tuning effect rather than acting as the dominant control parameter. Overall, the permutation analysis reveals a physically meaningful division of roles among the geometric variables: the structural period predominantly controls the resonance position, whereas the ITO thickness primarily determines the resonance strength and spectral confinement. These results provide additional physical insight into the surrogate model and support the selection of $P$ and $t_{\text{ITO}}$ as the principal variables governing the optical response of the proposed emitter. After establishing the surrogate models, Bayesian optimization (BO) was employed to determine new geometries with $E_{\text{peak}} \geq 0.90$, while simultaneously maintaining a narrowband solution via a penalty-weighted objective that penalizes wide bandwidth and out-of-band leakage. The convergence behavior of the Bayesian optimization procedure is presented in Fig. S2 of the Supplementary Materials.

### *3.1.4. Multi-metric screening of feasible Bayesian-optimization candidates*

Although all feasible Bayesian-optimization candidates satisfy the peak-emittance requirement ($E_{\text{peak}} \geq 0.90$), they still exhibit considerable differences in linewidth and spectral concentration. Consequently, selecting final candidates requires a multi-metric screening step that jointly considers peak location, linewidth, and sideband suppression. Fig. 6 (a) presents the distribution of the Bayesian optimization candidates in the predicted $\lambda_{\text{peak}}$–FWHM space, with the marker color representing the corresponding spectral-leakage metric. The result shows that candidates with similar peak wavelengths can exhibit substantially different linewidths, demonstrating that resonance position alone does not determine the spectral selectivity of the emitter. The distribution therefore provides a multi-metric view for identifying candidates that combine the desired resonance position with a relatively narrow emission bandwidth and reduced out-of-band contribution. A complementary view is provided in Fig. 6 (b), which plots all evaluated BO candidates in the $(\lambda_{\text{peak}}, E_{\text{peak}})$ plane and colors each point by the in-band fraction $f_{\text{in}}$, i.e., the fraction of total emission energy contained within the peak-centered in-band region. This highlights that designs with similar $E_{\text{peak}}$ may differ markedly in spectral concentration: higher $f_{\text{in}}$ corresponds to cleaner narrowband behavior with reduced spectral spread into the background. Overall, Figs. 6 (a) and (b) demonstrate that the feasible Bayesian-optimization candidates exhibit substantial variation in spectral quality despite satisfying the peak-emittance requirement. Accordingly, the final emitter cannot be selected on the basis of $E_{\text{peak}}$ alone; rather, peak wavelength, linewidth, and in-band spectral concentration must be considered simultaneously. This multi-metric screening provided the basis for selecting the most promising candidate for subsequent full-wave FDTD verification.

### *3.1.5. Optical performance of the optimized emitter*

Following multi-metric screening of the Bayesian optimization candidates, the selected geometry was directly verified using full-wave FDTD simulation. As shown in Fig. 6(c), the optimized emitter exhibits a near-unity peak emittance of $E_{\text{peak}} = 0.994$ at $\lambda_{\text{peak}} = 2181.96$ nm, with a FWHM of 169.29 nm, $f_{\text{in}} = 0.172$, and $R_{\text{side}} = 4.818$. The direct FDTD result therefore confirms that the surrogate-assisted optimization successfully identifies a geometry combining high peak emittance with a spectrally confined narrowband response. The optimized structure employs $t_{\text{ITO}} = 255.7$ nm, $t_{\text{Al}_2\text{O}_3} = 5.48$ nm and $P = 398nm$. To elucidate the physical origin of the near-unity resonant emittance, the spatial distributions of the electric- and magnetic-field intensities were evaluated at $\lambda_{\text{peak}} = 2181.96$ nm. As shown in Fig. 7 (a), the electric-field intensity, $|E|^2$, is strongly localized near the edges and interfaces of the patterned ITO layer, with pronounced hotspots appearing close to both the ITO/$Al_2O_3$ and ITO/sapphire boundaries. In contrast, the field intensity within the sapphire substrate remains comparatively weak. Since ITO exhibits appreciable optical loss in this spectral region, the localized electric field enhances the interaction between the resonant electromagnetic mode and the ITO layer, thereby promoting optical dissipation at the resonance wavelength. The numerical stability of the optimized optical response with respect to spatial mesh refinement is evaluated in Section 8 of the Supplementary Materials.

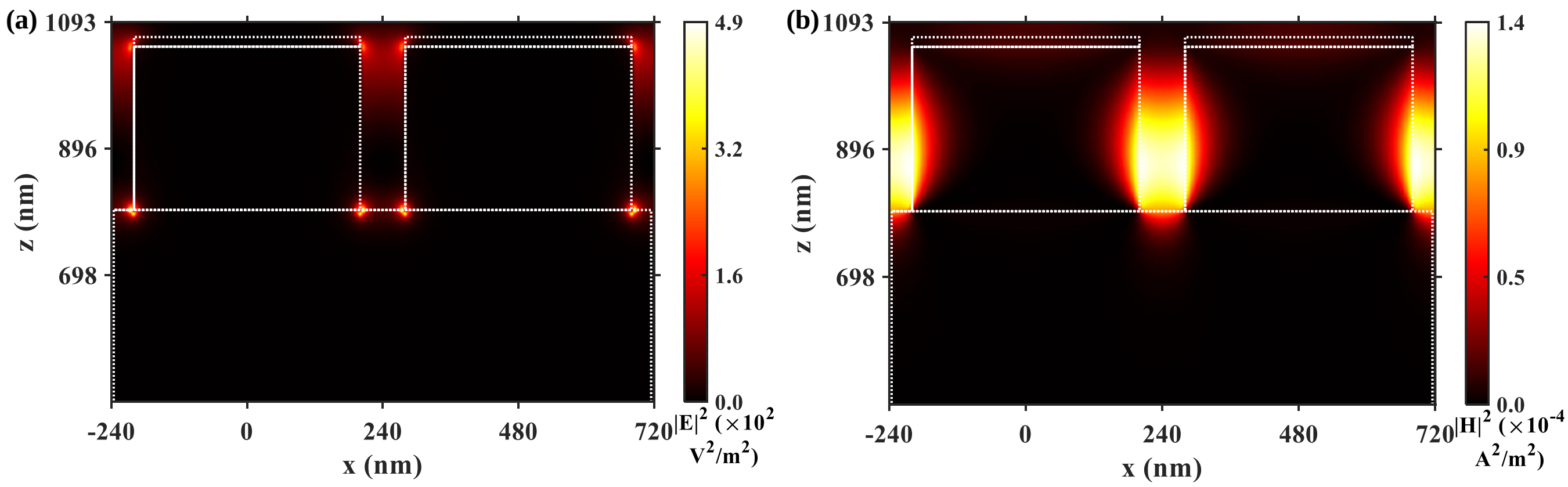

**Figure 7:** Spatial electromagnetic-field distributions of the optimized all-oxide emitter at the resonance wavelength $\lambda_{\text{peak}}$ = 2181.96 nm: (a) electric-field intensity $|E|^2$ and (b) magnetic-field intensity $|H|^2$. The white dotted outlines delineate the $Al_2O_3$, ITO, and sapphire regions. Pronounced field localization occurs within and near the patterned ITO region and its interfaces, consistent with resonantly enhanced electromagnetic confinement at the high-emittance wavelength.

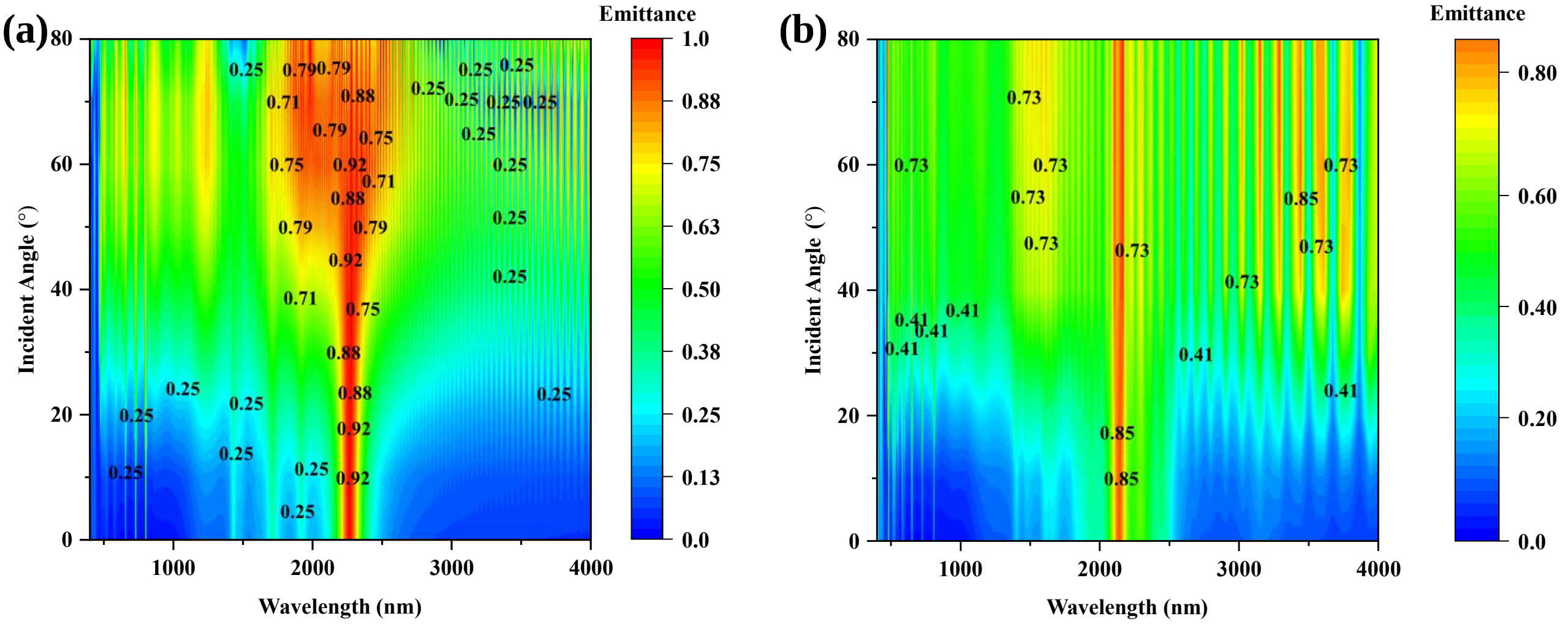

**Figure 8:** Angular response of the optimized all-oxide emitter for (a) TM and (b) TE polarizations.

The corresponding magnetic-field intensity, $|H|^2$, shown in Fig. 7(b), exhibits a broader confinement within the patterned ITO region, with particularly strong enhancement near the lower ITO interface and along the lateral boundaries of the periodic features. The complementary localization of the electric and magnetic fields indicates strong electromagnetic confinement supported by the optimized geometry. The weak penetration of both fields into the underlying sapphire further indicates that the dominant resonant interaction occurs primarily within the patterned $Al_2O_3$/ITO region. Consequently, the combination of geometry-induced field confinement and intrinsic optical loss in ITO produces enhanced resonant absorption, which, according to Kirchhoff's law, gives rise to the observed high thermal emittance.

Fig. 8 presents the angle-dependent spectral emittance of the optimized emitter for TM and TE polarizations over incident angles from 0° to 80°. According to Fig. 8 (a), for TM polarization, the dominant narrowband resonance remains concentrated near the designed 2.1–2.3 $\mu$m region and retains a high emittance of approximately 0.9–1.0 from normal incidence to about 50°. Within this angular range, only a limited displacement of the principal emission band is observed, indicating that the optimized resonance is comparatively insensitive to moderate changes in incidence angle. At larger angles, particularly above approximately 50°, the high-emittance region progressively broadens and separates into additional spectral branches, indicating a loss of the narrowband character obtained near normal incidence.

A similar trend can be observed for TE polarization in Fig. 8 (b). The principal emission feature remains identifiable near the optimized wavelength over low and moderate incidence angles, although its maximum emittance is somewhat lower than that obtained for TM polarization. The TE response also remains comparatively stable up to approximately 50°, beyond which additional wavelength-dependent features become increasingly pronounced and the spectral response becomes less confined to the original narrowband resonance. The asymmetrical 1D grating geometry of the structure may contribute to the difference between the spectral response at TM and TE polarization. Nevertheless, both polarizations preserve the principal emission band over a wide angular range.

## 3.2. Thermo-Optical and Electrothermal Analysis

The optimized emitter with $t_{Al_2O_3} = 5.48$ nm, $t_{ITO} = 255.7$ nm, and $P = 398$ nm exhibited the best narrowband performance among the evaluated candidate structures. We considered the emitter structure for further analysis to determine the electrothermal performance of the TPV cell. The TPV cell's electrical performance strongly depends on the emitter's emission spectrum, which changes with temperature because the emitter material's optical characteristics are temperature-dependent. For this reason, before proceeding with the electrothermal analysis, the complex refractive index of the ITO material at different temperatures was evaluated using temperature-dependent optical properties of ITO.

### 3.2.1. *Thermo-optical model of ITO*

The optical constants of ITO at elevated temperatures were obtained using a Drude–Lorentz model fitted to the room-temperature refractive index ($n$) and extinction coefficient ($k$) data reported by Del Villar *et al.* at 300 K [34]. Temperature dependence was subsequently introduced through the Drude damping parameter using the Bloch–Grüneisen relation.

Since no single published work provided both the necessary broad-wavelength optical constants and an appropriate analysis of their operating-temperature dependence, the optical data at 300K and the supporting physical models were taken from various studies. The extracted $n$ and $k$ data of Del Villar *et al.* were used as the reference optical parameters at 300K because they cover the wavelength range upto 4000 nm [34]. The Drude-Lorentz analysis presented by D'Elia *et al.* was utilized to substantiate the physical interpretation of the model parameters, including the association of the Drude term with the free-carrier response of ITO [35].

In the first step, the corresponding complex permittivity was calculated at 300K as follows:

$$\widetilde{\varepsilon}_{300\mathrm{K}}(\lambda, 300K) = \left[n_{300\mathrm{K}}(\lambda) + ik_{300\mathrm{k}}(\lambda)\right]^2, \qquad (31)$$

where the $n_{300\mathrm{K}}(\lambda)$ and $k_{300\mathrm{k}}(\lambda)$ denote extracted refractive index and extinction coefficient from Del Villar *et al.*, respectively. Equation (31) can be expressed as

$$\widetilde{\varepsilon}_{300\mathrm{K}}(\lambda, 300K) = \varepsilon_{1,300\mathrm{K}}(\lambda, 300K) + i\varepsilon_{2,300\mathrm{K}}(\lambda, 300K), \qquad (32)$$

with

$$\varepsilon_{1,300\mathrm{K}}(\lambda, 300K) = n^2_{300\mathrm{K}}(\lambda) - k^2_{300\mathrm{K}}(\lambda), \qquad (33)$$

$$\varepsilon_{2,300\mathrm{K}}(\lambda, 300K) = 2n_{300\mathrm{K}}(\lambda)k_{300\mathrm{K}}(\lambda). \qquad (34)$$

After that, a Drude-Lorentz model was fitted in the energy domain simultaneously to the extracted $n$ and $k$ values as follows:

$$\widetilde{\varepsilon}_{\mathrm{f}}(E) = \varepsilon_\infty - \frac{A_D}{E^2 + iB_D E} + \frac{A_L}{E_L^2 - E^2 - iB_L E}, \qquad (35)$$

where $E = \frac{hc}{\lambda}$ is the photon energy. $\varepsilon_\infty$ represents the high-frequency permittivity, $A_D$ is the Drude strength, $B_D$ denotes the Drude damping energy, and $A_L$, $B_L$, and $E_L$ are the strength, damping energy, and resonance energy of the Lorentz oscillator, respectively.

The $n$ and $k$ spectra extracted from Del Villar *et al.* contain a Drude term that represents the free-carrier contribution and the Lorentz term that indicates the bound-electron contribution. Therefore, they were fitted into the equation (35) to determine the Drude strength $A_D$ and damping energy $B_D(300)$ at 300K. Their reported Drude-Lorentz parameters were used as initial estimates for the nonlinear fitting process. Simultaneous fitting to the room temperature $n$ and $k$ spectra resulted in $\mathrm{RMSE}_n = 0.00206$ and $\mathrm{RMSE}_k = 0.00534$. The complete fitting procedure, fitted parameter values, and additional model details are provided in Section S6 and Table S4 of the Supplementary Material.

Temperature dependence was introduced through the Drude damping parameter $B_D$, which is associated with the free-carrier scattering rate. The Drude strength, high-frequency permittivity, and Lorentz-oscillator parameters were retained at their fitted 300 K values because independent temperature-resolved carrier-density, effective-mass, and interband-oscillator data for the same ITO film were not available. The temperature-dependent Drude damping was described using the normalized Bloch-Grüneisen relation

$$B_D(T) = B_D(300)\left[f_{\mathrm{res}} + (1 - f_{\mathrm{res}})\frac{\Phi_{\mathrm{BG}}(T, \Theta_D)}{\Phi_{\mathrm{BG}}(300, \Theta_D)}\right], \qquad (36)$$

where $f_{\mathrm{res}}$ represents the residual-scattering fraction and $\Theta_D$ is the effective transport Debye temperature. In the present model, $f_{\mathrm{res}} = 0.909$ and $\Theta_D = 1000$ K were obtained from the literature-informed transport analysis described in the section S5 of the Supplementary Material [36]. The Bloch–Grüneisen function is

$$\Phi_{BG}(T, \Theta_D) = \left(\frac{T}{\Theta_D}\right)^5 \int_0^{\Theta_D/T} \frac{x^5 e^x}{(e^x - 1)^2}\, dx. \quad (37)$$

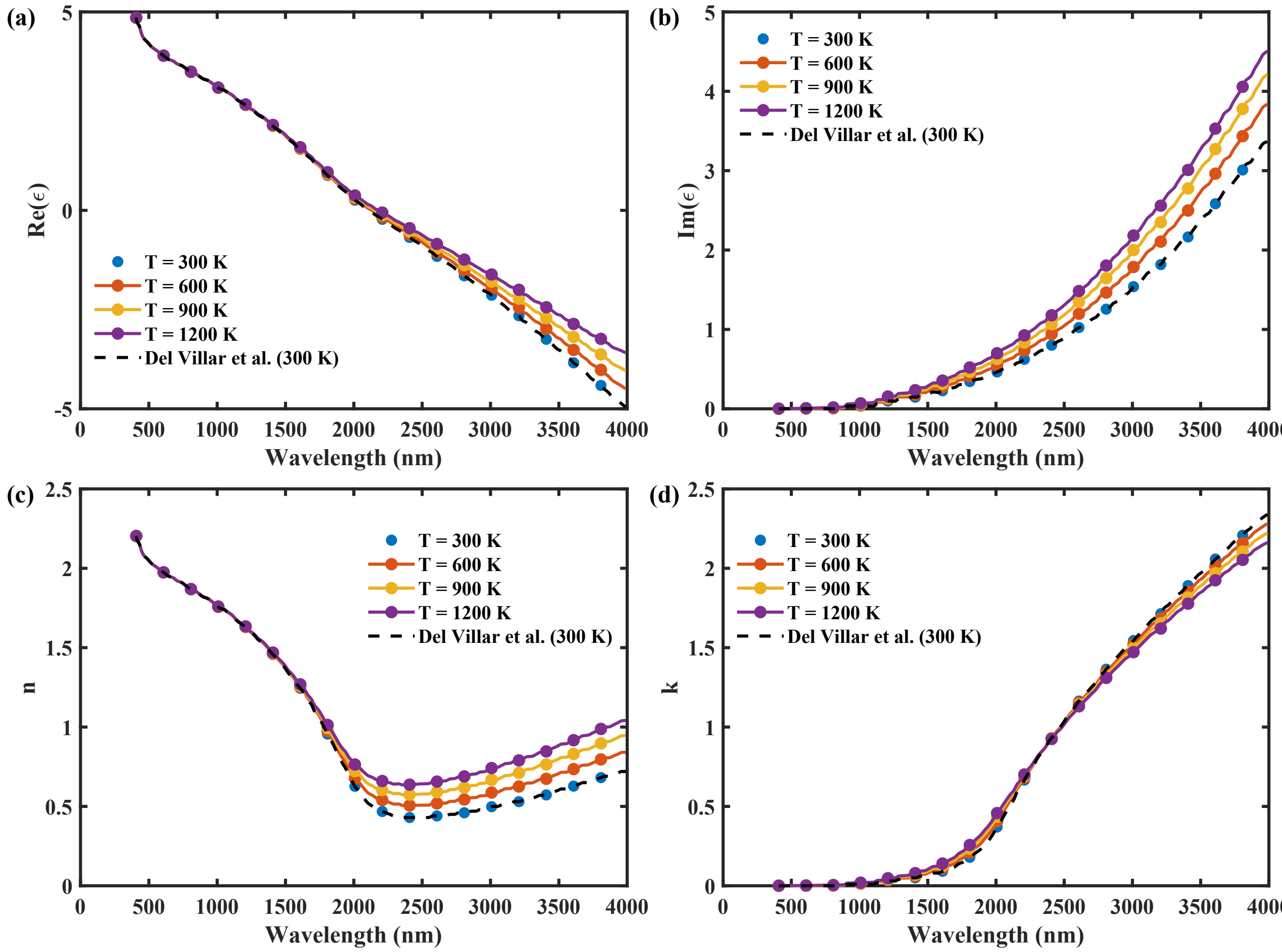

**Figure 9:** Temperature-dependent optical response of ITO obtained from the Bloch–Grüneisen-modified Drude–Lorentz model: (a) real part of the complex relative permittivity, Re($\varepsilon$); (b) imaginary part, Im($\varepsilon$); (c) refractive index $n$; and (d) extinction coefficient $k$ over 400–4000 nm at $T_{emi}$ = 300, 600, 900, and 1200 K. The black dashed curves denote the room-temperature optical data extracted from Del Villar *et al.*, demonstrating preservation of the reference response at 300 K.

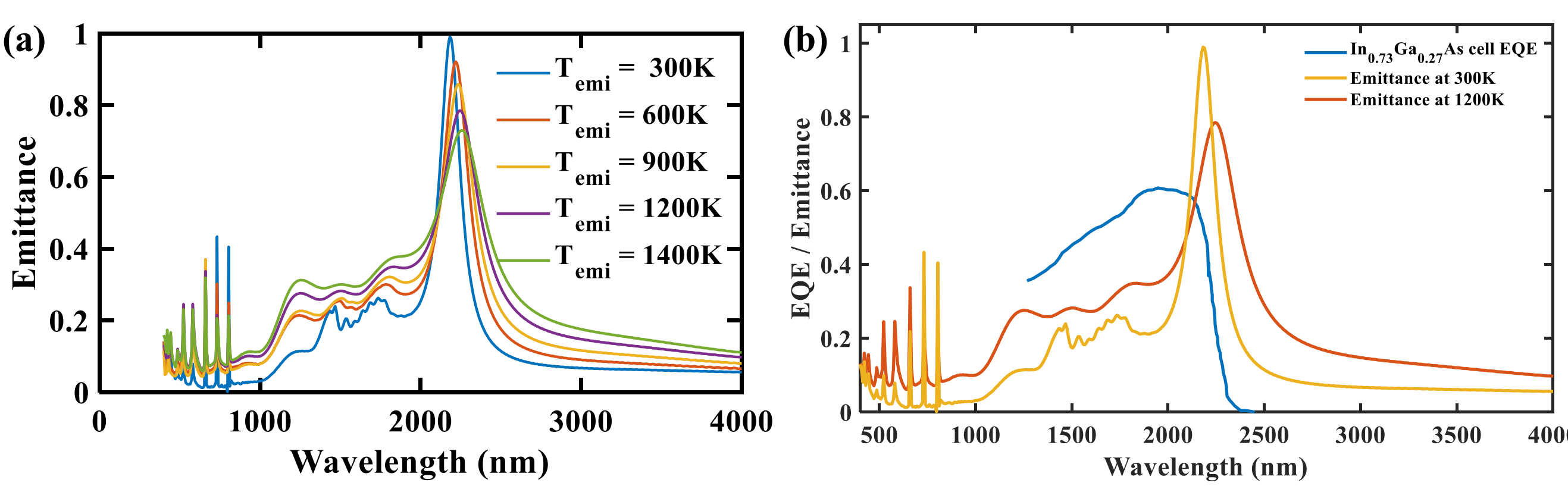

**Figure 10:** (a) Temperature-dependent emittance spectra of the optimized all-oxide emitter, (b) Spectral compatibility between the optimized emitter and the $In_{0.73}Ga_{0.27}As$ TPV cell, comparing the cell EQE with the emitter emittance at 300 and 1200 K.

At each temperature, the modified Drude contribution was calculated as

$$\widetilde{\varepsilon}_D(E,T) = -\frac{A_D(300)}{E^2 + iB_D(T)E}. \quad (38)$$

To preserve the experimentally referenced room-temperature optical response, the temperature-dependent complex permittivity was constructed as

$$\widetilde{\varepsilon}(E,T) = \widetilde{\varepsilon}_{\mathrm{DV}}(E,300) + \widetilde{\varepsilon}_D(E,T) - \widetilde{\varepsilon}_D(E,300), \quad (39)$$

where $\widetilde{\varepsilon}_{\mathrm{DV}}(E,300)$ denotes the complex permittivity obtained from the Del Villar room-temperature optical data. Equation (39) therefore exactly recovers the imported 300 K reference response while introducing only the temperature-dependent change in the Drude free-carrier contribution. The derivation of $f_{\mathrm{res}}$, details of the fitting procedure, and conversion of the resulting complex permittivity into $n(\lambda,T)$ and $k(\lambda,T)$ are provided in Section S5 of the Supplementary Material.

Figure 9 presents the calculated temperature-dependent dielectric response and optical constants of ITO over the wavelength range of 400–4000 nm. As shown in Fig. 9(a), the real part of the relative permittivity, Re($\varepsilon$), decreases continuously with wavelength and changes from positive to negative values near the epsilon-near-zero region. This behavior indicates a transition from a dielectric-like response to a free-carrier-dominated metallic response. With increasing temperature, Re($\varepsilon$) becomes slightly less negative in the infrared region because the temperature-induced increase in the Drude damping reduces the magnitude of the negative free-carrier contribution. Fig. 9(b) shows that the imaginary part of the relative permittivity, Im($\varepsilon$), increases with both wavelength and temperature, particularly beyond approximately 2000 nm. This trend is attributed to enhanced free-carrier scattering and the resulting increase in optical dissipation at elevated temperatures. The corresponding refractive index is presented in Fig. 9(c). The refractive index decreases sharply toward a minimum near the plasma-transition region and then increases gradually at longer wavelengths. The temperature-induced increase in $n$ becomes more pronounced in the infrared region, where the Drude free-carrier response dominates the optical behavior of ITO. Fig. 9(d) presents the extinction coefficient $k$. The extinction coefficient remains very small at shorter wavelengths and increases rapidly beyond the plasma-transition region. A slight decrease in $k$ is observed as temperature increases at longer wavelengths. Although Im($\varepsilon$) increases with temperature, this behavior is physically consistent because $k$ depends on both the real and imaginary parts of the complex permittivity. The simultaneous reduction in the magnitude of the negative Re($\varepsilon$) can therefore produce a moderate decrease in $k$. Overall, Fig. 9 demonstrates that the temperature dependence of the ITO optical response is comparatively weak in the visible and near-infrared regions but becomes increasingly significant in the mid-infrared region, where the free-carrier contribution is dominant. The close agreement between the 300 K results and the dashed room-temperature reference from Del Villar *et al.* [34] confirms that the original room-temperature optical response was preserved in the temperature-dependent material model.

Fig. 10 demonstrates the spectral response of the optimized all-oxide emitter structure at various temperatures. From Fig. 10, it can be seen that as the temperature increases, the principal emission band remains concentrated near the targeted (2.1–2.3 $\mu$m) region but gradually broadens, while its peak magnitude decreases and the off-resonant emission increases. This evolution is consistent with the temperature-dependent changes in the complex refractive index of ITO shown in Fig. 9. In particular, the increase in n in the infrared modifies the effective optical response of the ITO-containing structure, whereas the temperature-dependent variation of k alters the optical loss associated with the resonant mode. These changes originate from the increasing Drude damping, for which Fig. 9 shows a less negative Re($\varepsilon$) together with an increase in Im($\varepsilon$) at elevated temperatures. Consequently, the resonance becomes more strongly damped and less spectrally confined, leading to the progressive linewidth broadening and enhanced background emission observed in Fig. 10.

#### 3.2.2. *Electrothermal performance of the emitter-coupled TPV cell*

Fig. 10(b) illustrates the spectral compatibility between the optimized all-oxide emitter and the $In_{0.73}Ga_{0.27}As$ TPV cell. At room temperature, the optimized emitter exhibits a narrow, near-unity emission peak at $\lambda_{\mathrm{peak}} = 2181.96$ nm, located close to the photovoltaic bandgap cutoff wavelength of approximately $\lambda_g = 2198.6$ nm corresponding to $E_g = 0.564$ eV. Such spectral positioning is advantageous for TPV conversion because photons emitted immediately above the semiconductor bandgap can contribute to photocarrier generation while carrying relatively little excess energy, thereby limiting thermalization losses associated with higher-energy photons. At a higher emitter operating temperature of $T_{\mathrm{emi}} = 1200$ K, the temperature-dependent optical response of ITO causes the principal emission band to broaden and shift toward longer wavelengths. Nevertheless, from Fig. 10(b), it can be observed that a substantial portion of the elevated-temperature emission remains within the photoresponsive spectral range of the TPV cell. The spectral broadening can increase the cell temperature by enhancing carrier thermalization at shorter wavelengths.

The temperature-dependent electrothermal analysis was first performed without heat sink configuration shown in Fig. 11(a) to determine the intrinsic temperature-dependent response of the emitter-coupled TPV cell. No external heat sink was included in this configuration. After identifying the emitter temperature corresponding to the maximum conversion efficiency, an additional simulation was performed at this operating point using the configuration shown in Fig. 11(b). In this case, a Cu heat sink was incorporated beneath the photovoltaic cell, and forced-air convection with $h = 75\ \mathrm{W\,m^{-2}K^{-1}}$ and $T_\infty = 300$ K was applied to the exposed outer surfaces of the heat sink. The TPV cell comprises $n^{++}$-$In_{0.73}Ga_{0.27}As$, $p$-$In_{0.73}Ga_{0.27}As$, and $p^{++}$-$In_{0.73}Ga_{0.27}As$ semiconductor layers together with Ag front and Al back contacts. Incident thermal photons are absorbed within the photovoltaic structure, contributing to photocarrier generation and electrical power conversion. A Cu heat sink is integrated beneath the cell to dissipate excess heat, and the outward radiation pattern indicates heat dissipation to the surroundings. The corresponding device and electrothermal simulation parameters, along with the cell materials parameters, are provided in Section S6 of the Supplementary Materials.

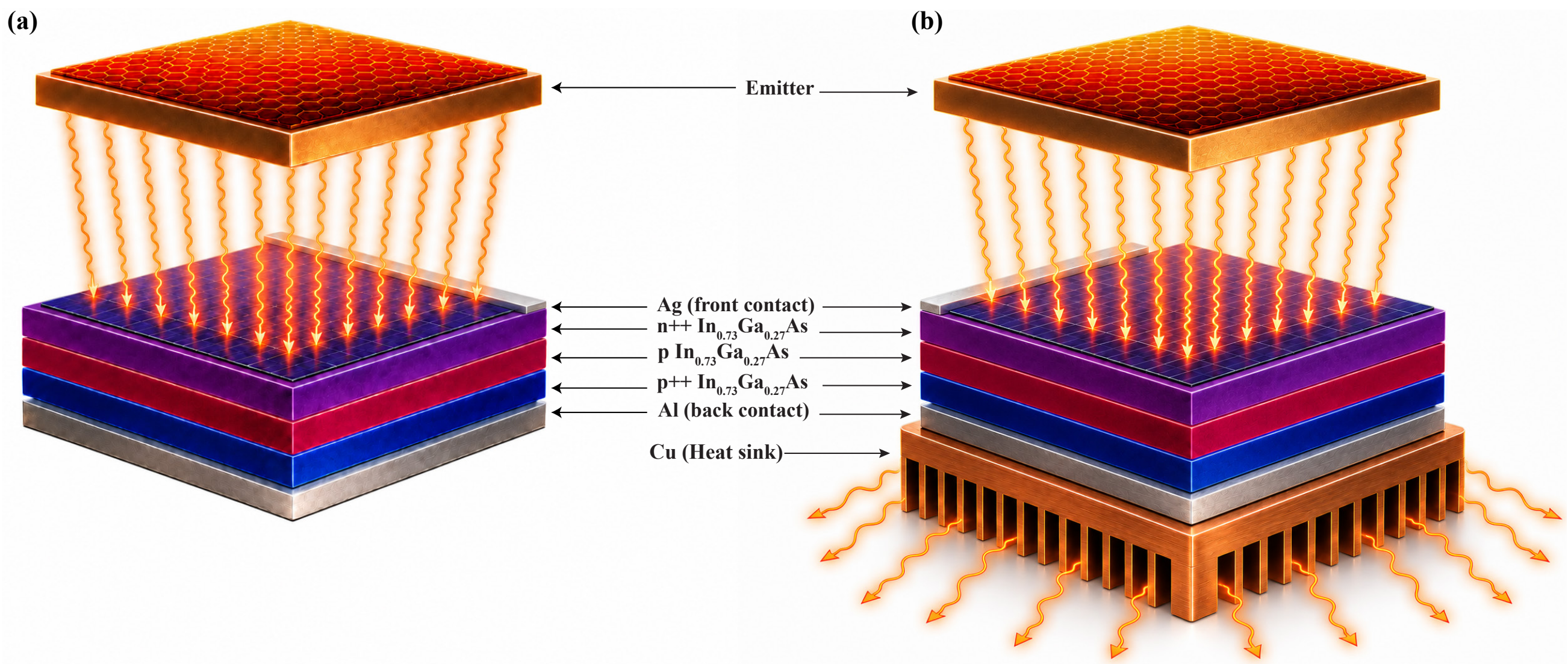

**Figure 11:** Schematic of the emitter-coupled $In_{0.73}Ga_{0.27}As$ TPV system used for the electrothermal analysis: (a) without any heat sink, in which heat is dissipated directly from the exposed TPV-cell surface to the surrounding air by convection without an external heat sink, and (b) active-cooling configuration, in which a Cu heat sink is incorporated beneath the TPV cell and forced-air convection is applied to the exposed heat-sink surfaces with $h = 75\ \mathrm{W\,m^{-2}K^{-1}}$ and $T_\infty = 300$ K.

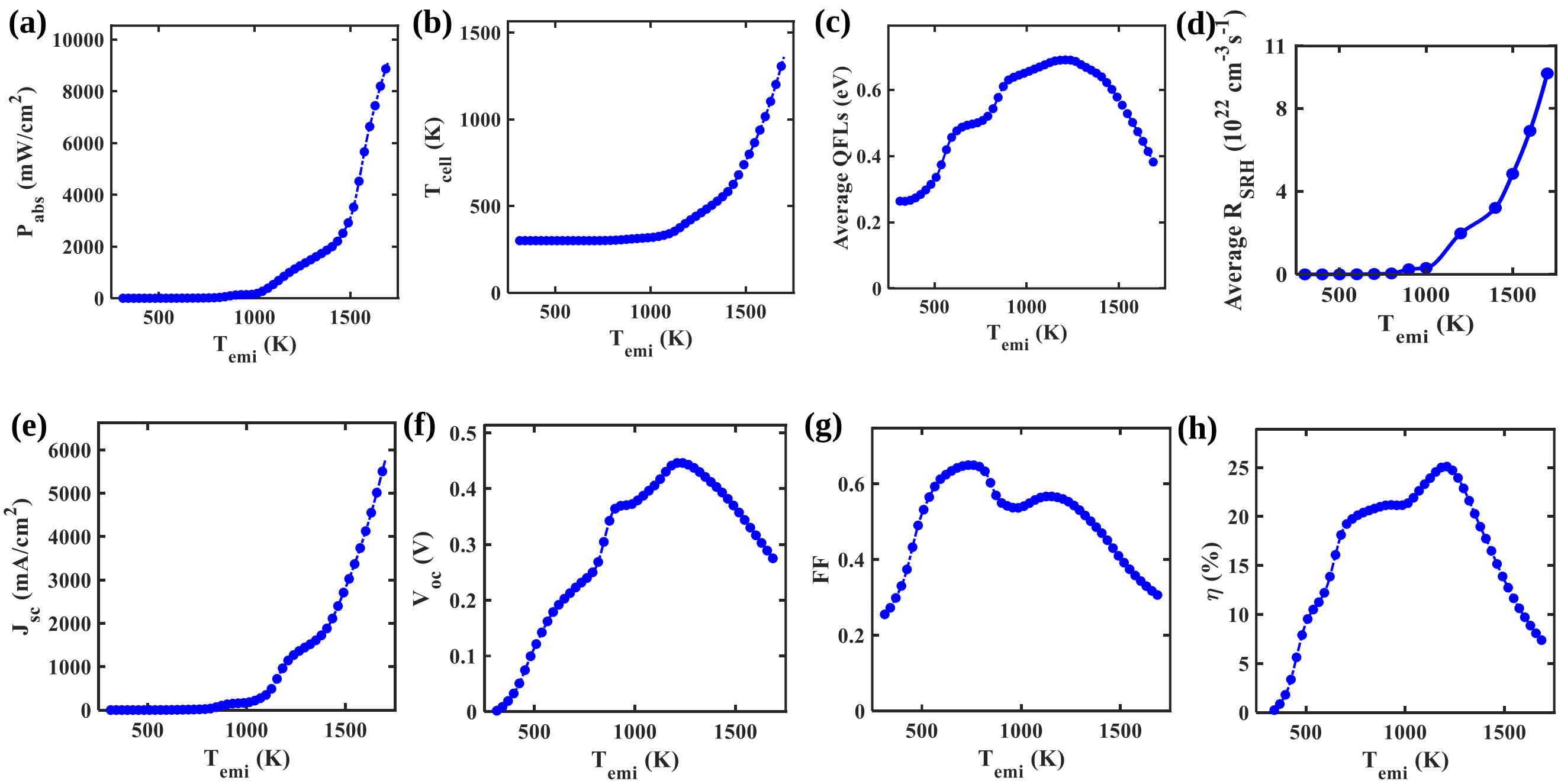

**Figure 12:** Temperature-dependent electrothermal response of the optimized-emitter-coupled $In_{0.73}Ga_{0.27}As$ TPV cell without heat sink configuration: (a) incident radiative power density $P_{in}$, (b) cell temperature $T_{cell}$, (c) spatially averaged quasi-Fermi-level splitting (QFLS) over the active semiconductor region, (d) spatially averaged Shockley-Read-Hall recombination rate $R_{SRH}$, (e) short-circuit current density $J_{sc}$, (f) open-circuit voltage $V_{oc}$, (g) fill factor (FF), and (h) TPV conversion efficiency $\eta_{TPV}$, as functions of emitter temperature $T_{emi}$.

Fig. 12 presents the temperature-dependent electrothermal response of the optimized-emitter-coupled $In_{0.73}Ga_{0.27}As$ TPV cell without a heat sink configuration. As shown in Fig. 12 (a), the absorbed radiative power density by the cell, $P_{abs}$, increases rapidly with emitter temperature, $T_{emi}$, due to the strong temperature dependence of the thermal photon flux. The increasing radiative loading consequently produces substantial self-heating of the photovoltaic device. Accordingly, Fig. 12 (b) shows that the cell temperature, $T_{cell}$, remains close to the ambient temperature at relatively low $T_{emi}$, but rises increasingly rapidly at elevated emitter temperatures as the absorbed thermal power increases. The resulting temperature dependence of the internal carrier processes is illustrated in Fig. 12(c) and Fig. 12(d). The spatially averaged quasi-Fermi-level splitting (QFLS) initially increases with $T_{emi}$, reflecting the increasing photogenerated carrier population under stronger above-bandgap illumination. The QFLS reaches a maximum in the intermediate-temperature region and subsequently decreases at higher $T_{emi}$, where cell heating increasingly enhances carrier losses. This interpretation is supported by Fig. 12(d), which shows a pronounced increase in the spatially averaged Shockley–Read–Hall recombination rate, $R_{SRH}$, at elevated emitter temperatures. Thus, the beneficial effect of increased photogeneration is progressively counteracted by temperature-induced recombination as the cell warms. The corresponding electrical characteristics are shown in Fig. 12(e)–(h). The short-circuit current density, $J_{sc}$, increases continuously with $T_{emi}$, as shown in Fig. 12(e), because the increasing above-bandgap photon flux produces a progressively larger photocarrier-generation rate. In contrast, the open-circuit voltage shown in Fig. 12(f) exhibits a non-monotonic response. The current–voltage characteristics were obtained from the carrier-transport formulation described by equations 13 and 14, together with the Poisson and carrier-continuity equations, and $V_{oc}$ was determined from the open-circuit condition at which the net terminal current density becomes zero. The $V_{oc}$ initially increases as the enhanced illumination raises the photogenerated carrier population and quasi-Fermi-level splitting. It reaches a maximum near the intermediate-temperature region and subsequently decreases as the increasing $T_{cell}$ enhances carrier recombination and reduces the achievable quasi-Fermi-level separation.

A similar competition is reflected in the fill factor shown in Fig. 12(g). The FF initially improves as the photovoltaic response strengthens with increasing illumination but then decreases at elevated emitter temperatures, where stronger cell heating and recombination progressively degrade the shape of the $J$–$V$ characteristic. Consequently, the conversion efficiency shown in Fig. 12(h), evaluated using the absorbed radiative power according to equation 22, initially increases with emitter temperature and reaches its maximum near $T_{emi} \approx 1200$ K. Beyond this operating point, the continued increase in $J_{sc}$ is insufficient to compensate for the simultaneous reductions in $V_{oc}$ and FF and the increasing thermal loading of the cell. The resulting reduction in efficiency at higher emitter temperatures demonstrates the competition between enhanced photon-driven carrier generation and temperature-induced recombination and thermal losses, identifying approximately 1200 K as the preferred emitter operating temperature.

Fig. 13 presents the temperature-dependent electrical output characteristics of the optimized-emitter-coupled $In_{0.73}Ga_{0.27}As$ TPV cell. As shown in Fig. 13(a), the short-circuit current density increases substantially with increasing emitter temperature, $T_{emi}$. This behavior originates from the rapid increase in thermal photon flux emitted by the selective emitter at elevated temperatures. The resulting increase in above-bandgap photon absorption enhances the photocarrier-generation rate within the $In_{0.73}Ga_{0.27}As$ cell and consequently increases $J_{sc}$. The voltage dependence, however, exhibits a different trend. The open-circuit voltage initially increases as the stronger illumination enhances photogenerated carrier density and quasi-Fermi-level splitting. However, at higher emitter temperatures, the corresponding rise in cell temperature enhances the intrinsic carrier concentration, which intensifies carrier recombination and consequently reduces the achievable quasi-Fermi-level separation. As a result, $V_{oc}$ begins to decrease despite the continued increase in photogenerated current. Consequently, the $J$–$V$ curves evolve toward considerably higher current densities but reduced open-circuit voltages at the highest emitter temperatures. The corresponding power-density characteristics are shown in Fig. 13(b), where the electrical power density is obtained from $P(V) = J(V)V$. For each emitter temperature, the power density increases from zero at short-circuit conditions, reaches a maximum at an intermediate operating voltage, and subsequently decreases to zero as the voltage approaches $V_{oc}$. The maximum of each curve, therefore, represents the maximum-power operating point of the TPV cell. Increasing $T_{emi}$ substantially enhances the maximum obtainable power density because of the strong increase in $J_{sc}$. However, the $V_{oc}$ decreases beyond temperature around $T_{emi} \approx 1200$ K, which leads to a decrease in efficiency beyond that temperature according to equation 22, despite an increase $J_{sc}$.

Fig. 14 compares the electrical output characteristics of the optimized-emitter-coupled TPV cell without heat sink and with heat sink configurations at $T_{emi} = 1200$ K. Without heat sink, the cell temperature reaches approximately 413 K, whereas the implementation of active thermal management reduces $T_{cell}$ to approximately 310 K. As shown in Fig. 14(a), the short-circuit current densities under the two thermal conditions remain nearly identical because the incident radiative spectrum and the resulting photogeneration are essentially unchanged. However, the actively cooled cell maintains a higher current density over a broader voltage range and exhibits a pronounced increase in the open-circuit voltage from approximately 0.44 V to about 0.50 V with active cooling from the heat sink. The substantial reduction in cell temperature mitigates temperature-induced carrier-recombination losses and helps preserve a larger quasi-Fermi-level separation, thereby improving the open-circuit voltage and the overall shape of the $J$–$V$ characteristic. The increased squareness of the actively cooled $J$–$V$ curve

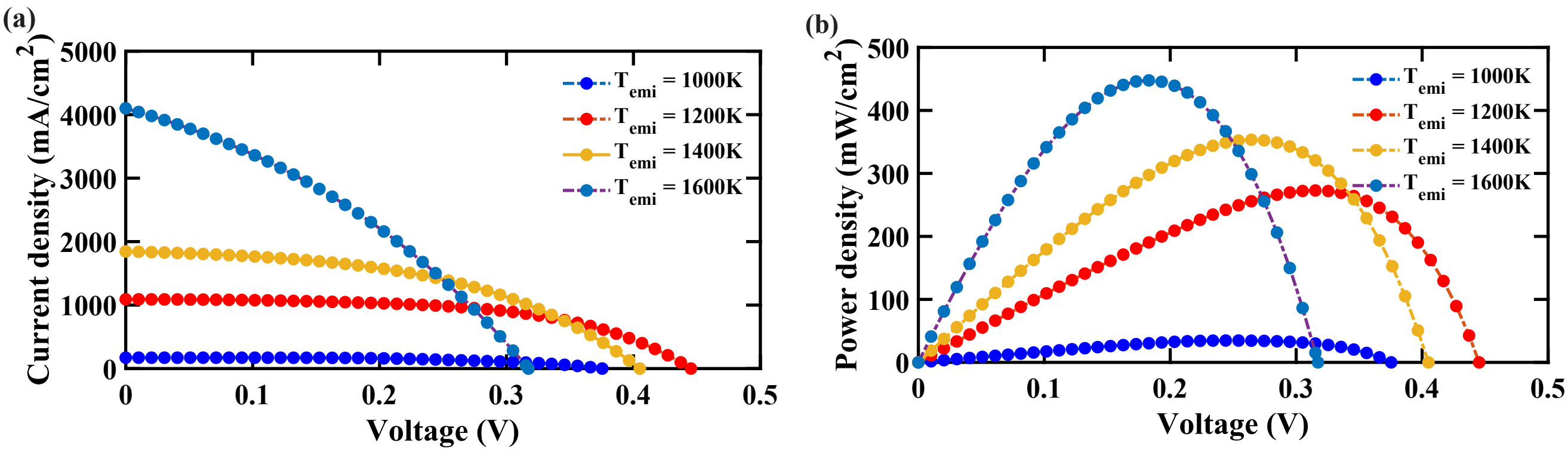

**Figure 13:** Temperature-dependent (a) current-density vs voltage ($J$–$V$) and (b) power-density vs voltage ($P$–$V$) characteristics at $T_{emi}$ = 1000, 1200, 1400, and 1600 K of the optimized-emitter-coupled $In_{0.73}Ga_{0.27}As$ TPV cell without heat sink configuration

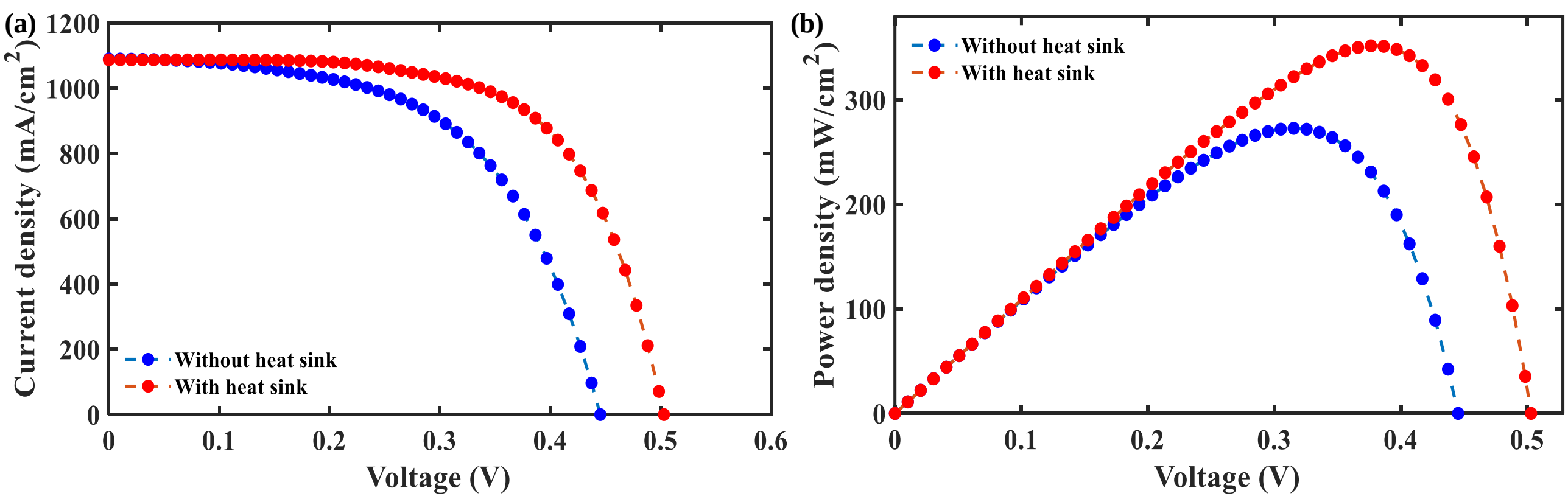

**Figure 14:** (a) current-density vs voltage ($J$–$V$) and (b) power-density vs voltage ($P$–$V$) characteristics of the TPV cell without heat sink and with heat sink configurations

further indicates an improvement in the fill factor. The corresponding power-density characteristics are presented in Fig. 14(b). Without a heat sink, the maximum power density is approximately 275 mW cm$^{-2}$, whereas active cooling with a heat sink increases the maximum electrical power density to approximately 350 mW cm$^{-2}$. The maximum-power operating point also shifts toward a higher voltage following the reduction in cell temperature. Since $J_{sc}$ changes only marginally between the two cases, the enhancement in electrical output primarily originates from the recovery of $V_{oc}$ and the fill factor rather than from an increase in photogenerated current. These results demonstrate that reducing $T_{cell}$ from approximately 413 K to 310 K through active thermal management substantially suppresses temperature-induced electrical losses and enables greater power extraction from the TPV cell at the same emitter operating temperature.

### 3.3. Comparative analysis

Table 1 compares the proposed all-oxide selective emitter with reported thermal-emitter and TPV architectures. Silva-Oelker *et al.* investigated $W/HfO_2/W$ and $Mo/HfO_2/Mo$ disk-array emitters coupled to GaSb cells in the far field. Their reported efficiencies were 20.99% and 20.38%, respectively, at 1685 K, with high emittance maintained over a relatively broad angular range [37]. The $W/Al_2O_3$ multilayer Fabry–Pérot emitter reported by Zhao *et al.* was coupled to a GaInAsSb cell and achieved a TPV efficiency of 22.59% at 1600 K while maintaining a strong angular response [38]. Li *et al.* reported a $Ta/SiO_2$ cross-patch metamaterial emitter with a narrow emission peak near 1.70 $\mu$m and when the emitter was coupled to InGaAsSb and Si/InGaAsSb tandem photovoltaic configurations, efficiencies of 41.68% and 46.26%, respectively, were reported at 1700 K [39]. These higher efficiencies, however, correspond to different photovoltaic architectures and operating conditions from those considered in the present work. In contrast, Rong *et al.*, focused on an emitter-only $Ge/SiO_2/Au$ nanocylinder configuration exhibiting an extremely narrow spectral response and high directionality [40]. Since that study did not include photovoltaic conversion analysis, its performance is more appropriately compared in terms of spectral selectivity and angular behavior rather than electrical efficiency.

However, the present $Al_2O_3$/ITO/sapphire emitter exhibits a near-unity peak emittance of $E_{peak}$ = 0.994 at $\lambda_{peak}$ = 2.182 $\mu$m with a FWHM of 169.29 nm, while the principal emission mode is maintained up to approximately 50°. When coupled to an $In_{0.73}Ga_{0.27}As$ TPV cell, the system reaches a maximum conversion efficiency of 25.12% at $T_{emi} = 1200$ K without an external heat sink. At the same emitter temperature, incorporation of the Cu heat sink reduces the cell temperature and increases the conversion efficiency to approximately 32.17%. Although the proposed configuration does not provide the highest conversion efficiency among the systems listed in Table 1, its distinguishing feature is the integration of an all-oxide, oxidation-resistant emitter architecture with surrogate-assisted Bayesian inverse design, direct FDTD verification, temperature-dependent ITO optical modeling, angular-response analysis, and cell-level electrothermal simulation within a unified computational framework. In addition, the optimum operating temperature of 1200 K is lower than the refractory-metal-based TPV systems listed in Table 1.

**Table 1**
Comparison of the proposed all-oxide selective emitter with representative thermal emitters reported in the literature.

| Emitter structure | PV cell / regime | Spectral performance | Angular response | Operating temperature | Reported efficiency | Reference |
|---|---|---|---|---|---|---|
| W/$HfO_2$/W and Mo/$HfO_2$/Mo disk arrays | GaSb Far-field | High emittance below GaSb cutoff $\lambda_g \approx 1.72$ $\mu$m | High emittance maintained to $\sim 60°$ | 1685 K | 20.99% (W) 20.38% (Mo) | [37] |
| W/$Al_2O_3$ multilayer Fabry–Pérot | GaInAsSb Far-field | Average emittance = 0.961 0.75–2.54 $\mu$m Peak > 0.99 | > 0.94 at 60° (TM) > 0.91 at 50° (TE) | 1600 K | 22.59% | [38] |
| Ta/$SiO_2$ cross-patch metamaterial | InGaAsSb and Si/InGaAsSb Far-field | $E_{peak} = 0.9998$ $\lambda_{peak} \approx 1.70$ $\mu$m | Stable to $\sim 45°$ Polarization insensitive | 1700 K | 41.68% InGaAsSb 46.26% tandem | [39] |
| Ge/$SiO_2$/Au nanocylinder array | Emitter-only optical study | $\lambda_{peak} = 3.222$ $\mu$m $E_{peak} > 0.97$ $Q = 536.7$ (TE) $Q = 402.5$ (TM) | < 0.2° (TE) $\sim 1°$ (TM) divergence | – | – | [40] |
| $Al_2O_3$/ITO/ sapphire all-oxide emitter | $In_{0.73}Ga_{0.27}As$ Far-field | $E_{peak} = 0.99$ $\lambda_{peak} = 2.182$ $\mu$m FWHM = 169 nm | Principal mode maintained to $\sim 50°$ | 1200 K optimum | 25.12% without heat sink $\sim 32.17\%$ with heat sink | This work |

### 3.4. Fabrication feasibility and proposed process

The optimized $Al_2O_3$/ITO/sapphire emitter can be realized using thin-film deposition and nanofabrication techniques. Firstly, the ITO structure can be fabricated using lithographic patterning and magnetron-sputtering techniques [28, 41]. In the first step of fabrication, the sapphire substrate can be cleaned using standard solvents and deionized water. Radio-frequency (RF) magnetron sputtering can then be used to deposit the ITO layer with a thickness of approximately 255.7 nm [29, 41]. Following ITO deposition, an $Al_2O_3$ layer of around 5.48 nm can be applied using atomic-layer deposition (ALD). The self-regulating surface-reaction mechanism of ALD facilitates the precise deposition of ultrathin $Al_2O_3$ films, where trimethylaluminum and water can be utilized as the metal and oxygen precursors, respectively, for ALD-grown $Al_2O_3$ [42]. Depositing the ITO and $Al_2O_3$ layers prior to lithographic patterning may be advantageous for the proposed configuration, as it allows the subsequently patterned features to retain the $Al_2O_3$ top layer used in the numerical model. The 1D periodic pattern can be defined using electron-beam lithography (EBL) for small-scale realization. Nanoimprint or similar scalable lithographic techniques may offer a more reliable path for large-area manufacturing. In particular, Hasuike *et al.* used thermal nanoimprint lithography in conjunction with RF sputtering to experimentally demonstrate an ITO diffraction-grating structure [41], while Dao *et al.* demonstrated scalable colloidal lithography for patterned ITO-based selective thermal emitters [28].

Following lithography, a sequential dry-etch procedure can be used to transfer the design into the $Al_2O_3$/ITO stack. A short and precisely controlled inductively coupled plasma reactive-ion etching (ICP–RIE) process based on $BCl_3/Cl_2$/Ar can be used to etch the $Al_2O_3$ layer since the $Al_2O_3$ cap is only about 5.48 nm thick. Anisotropic dry etching of $Al_2O_3$ thin films has been experimentally demonstrated using such plasma chemistries [43]. After that, a chlorine-containing plasma can be used to etch the underlying ITO. Yoon *et al.* provided an experimental foundation for anisotropic pattern transfer in ITO by demonstrating controlled dry etching of ITO-containing multilayer structures using $H_2$/HCl-based high-density plasma [44]. To maintain the optimal ITO thickness and lateral structure, the etch time and over-etch should be reduced.

After pattern transfer, the remaining lithographic mask can be removed, and the manufactured geometry can be examined using atomic force microscopy and scanning electron microscopy to ascertain the grating periodicity, lateral dimensions, and surface morphology. The ITO and $Al_2O_3$ thicknesses can also be confirmed using calibrated profilometry or cross-sectional electron microscopy. To find the experimentally realized complex refractive index, spectroscopic ellipsometry should be used on a blanket ITO witness film that has undergone the same processing. Because the free-carrier optical response of ITO is sensitive to thermal processing, crystallinity, and oxygen partial pressure, this step is especially crucial for the current emitter [30, 35].

## 4. Conclusion

In this paper, a surrogate-assisted inverse-design framework was developed for an all-oxide narrowband thermophotovoltaic emitter comprising a patterned $Al_2O_3$/ITO structure on a sapphire substrate. The narrowband emission FOMs of the dataset spectra obtained from FDTD analysis were subsequently used to train ExtraTrees surrogate models. The trained models reproduced the major design-space trends with strong predictive performance for FWHM and $E_{\text{peak}}$ while moderate performance for $\lambda_{\text{peak}}$ and $f_{\text{in}}$. After determining the optimized structure from surrogate-assisted Bayesian optimization, the final selected geometry was verified directly using full-wave FDTD simulation. The optimized emitter achieved a near-unity peak emittance of $E_{\text{peak}} = 0.994$ at $\lambda_{\text{peak}} = 2181.96$ nm with an FWHM of 169.29 nm. Angular analysis further revealed that the peak-emittance remains prominent over a broad angular range, with more noticeable broadening and mode splitting occurring above approximately 50°. The obtained electric-field distribution at the resonance wavelength exhibited strong field localization near the patterned interfaces, supporting the resonant origin of the enhanced optical response. The temperature-dependent optical performance analysis demonstrated that increasing temperature escalates free-carrier damping, leading to progressive broadening of the peak emission band. The electrothermal performance of the optimized emitter was assessed for a $In_{0.73}Ga_{0.27}As$ cell whose EQE is compatible with the emission spectrum of the emitter. The electrothermal analysis exhibited that increasing emitter temperature enhances photogeneration, and eventually, the cell temperature, which progressively increases dark current losses and recombination. Consequently, the conversion efficiency increased with emitter temperature and reached a maximum of approximately 32.17% near $T_{\text{emi}} = 1200$ K with the heat sink configuration considered in this work. Overall, the obtained results demonstrate that surrogate-assisted Bayesian inverse design of $Al_2O_3$/ITO/sapphire architecture with temperature-dependent material modeling and electrothermal analysis can provide an effective computational route for developing spectrally selective TPV emitters in an oxidation-resistant material platform for high-temperature operation.

## CRediT authorship contribution statement

**Bibekananda Nath:** Conceptualization, Methodology, Visualization, Software, Investigation, Writing – original draft. **Kawshik Nath:** Methodology, Visualization, Software, Investigation, Writing – original draft. **Ahmed Zubair:** Conceptualization, Methodology, Visualization, Resources, Writing – original draft, Writing – review & editing, Supervision.

## Data Availability Statement

The data supporting the findings presented in this paper are not currently available to the public, but they may be obtained from the authors upon reasonable request.

## Acknowledgements

The authors thank the Bangladesh University of Engineering and Technology (BUET) for providing technical support and computational facilities.

## References

[1] RJ Nicholas and RS Tuley. Thermophotovoltaic (TPV) devices: Introduction and modelling. In *Functional materials for sustainable energy applications*, pages 67–90. Elsevier, 2012. doi: 10.1533/9780857096371.1.67.

[2] Tao Cui, Yan Shen, Ao Cheng, Zhe Liu, Shi Jia, Shuai Tang, Lei Shao, Huanjun Chen, and Shaozhi Deng. Highly efficient molybdenum nanostructures for solar thermophotovoltaic systems: One-step fabrication of absorber and design of selective emitter. *Chemical Engineering Journal*, 487:150389, 2024. doi: 10.1016/j.cej.2024.150389.

[3] Bibekananda Nath and Ahmed Zubair. Broadband high-temperature multilayer pyramid-shaped metamaterial thermal absorber for thermophotovoltaic applications. *Case Studies in Thermal Engineering*, 78:107697, 2026. ISSN 2214-157X. doi: https://doi.org/10.1016/j.csite.2026.107697.

[4] Yili Tang, Zhuming Liu, Ximeng Chen, Jiapeng Li, Yonghui Liu, Xiaoyu Lv, Xincun Peng, Liangliang Tang, and Jianxiong Shao. Combining of anodic oxidization with Zn-Ga diffusion to fabricate high-efficiency GaSb thermophotovoltaic cells. *IEEE Transactions on Electron Devices*, 71(4):2585–2591, 2024. doi: 10.1109/TED.2024.3362311.

[5] Govind Padmakumar, Aravind Balaji, Federica Saitta, Paula Perez-Rodriguez, René A.C.M.M. van Swaaij, and Arno H.M. Smets. Hexagonal microtextured glass to achieve high optical performance in thin-film silicon solar cells. *Solar Energy*, 306:114292, 2026. ISSN 0038-092X. doi: 10.1016/j.solener.2025.114292.

[6] Wenbin Lin, Yu Cao, Zhicheng Ke, and Ali Hassan. Shape-driven optimization strategy for efficient and stable lead-free all-perovskite tandem solar cells. *Solar Energy Materials and Solar Cells*, 295:113971, 2026. ISSN 0927-0248. doi: 10.1016/j.solmat.2025.113971.

[7] Yeonhwa Kim, Hyun-Beom Shin, Eunkyo Ju, Tsimafei Laryn, Taehee Kim, In-Hwan Lee, Ho Kwan Kang, Won Jun Choi, and Daehwan Jung. Enhanced short-circuit current density in epitaxial InGap/GaAs/Si triple-junction solar cells enabled by wide bandgap n-AlGaAs buffers. *Solar Energy Materials and Solar Cells*, 297:114133, 2026. ISSN 0927-0248. doi: 10.1016/j.solmat.2025.114133.

[8] Vishnu Narayanan V, K.S. Rajni, Ipsita Jena, and Udai P. Singh. Efficiency enhancement of CdTe solar cells using $Cu_2MnSnS_4$ back contact layer: - experimental and numerical analysis. *Renewable Energy*, 260:125183, 2026. ISSN 0960-1481. doi: 10.1016/j.renene.2026.125183.

[9] Qiong Peng, Sahibzada Muhammad Zaheer, Jingfeng Li, Javed Rehman, Saiful Arifin Shafiee, M. Kashif Masood, and Norah Salem Alsaiari. Enhancing optical absorption and efficiency of perovskite solar cells using embedded ag nanoparticles array and ito moth-eye anti-reflective layers. *Optical and Quantum Electronics*, 58(2):69, Jan 2026. ISSN 1572-817X. doi: 10.1007/s11082-025-08633-y.

[10] Leiping Duan, Xin Cui, Cheng Xu, Zhong Chen, and Jianghui Zheng. Monolithic Perovskite/Perovskite/Silicon Triple-Junction Solar cells: Fundamentals, Progress, and Prospects. *Nano-Micro Letters*, 18(1): 8, Jul 2025. ISSN 2150-5551. doi: 10.1007/s40820-025-01836-8.

[11] Manisha Rautela, Sumit Sagar, and Jitendra Kumar. Hourglass-shaped $GaAs_{0.99}Bi_{0.01}$ nanowire solar cells with CuI-PEDOT:PSS double hole transport layers for enhanced photovoltaic performance. *Scientific Reports*, Jan 2026. ISSN 2045-2322. doi: 10.1038/s41598-025-34717-6.

[12] Sukanta Dhar, Sourav Mandal, Gourab Das, Sampad Mukherjee, Chandan Banerjee, H. Saha, and A. K. Barua. ITO nanorod-driven enhancement of current density in a-Si based p-i-n solar cells. *Journal of Materials Science*, 61(7):4526–4543, Feb 2026. ISSN 1573-4803. doi: 10.1007/s10853-025-12071-2.

[13] Md Faiaad Rahman, Md Ashaduzzaman Niloy, Ehsanur Rahman, and Ahmed Zubair. Unveiling architectural and optoelectronic synergies in lead-free perovskite/perovskite/kesterite triple-junction monolithic tandem solar cells. *arXiv preprint arXiv:2511.06059*, 2025. doi: 10.48550/arXiv.2511.06059.

[14] Ian Marius Peters, Carlos David Rodríguez Gallegos, Larry Lüer, Jens A Hauch, and Christoph J Brabec. Practical limits of multijunction solar cells. *Progress in Photovoltaics: Research and Applications*, 31(10):1006–1015, 2023. doi: doi.org/10.1002/pip.3705.

[15] Yun Da and Yimin Xuan. Role of surface recombination in affecting the efficiency of nanostructured thin-film solar cells. *Optics express*, 21(S6):A1065–A1077, 2013. doi: 10.1364/OE.21.0A1065.

[16] Etienne Moulin, Ulrich Wilhelm Paetzold, Hilde Siekmann, Janine Worbs, Andreas Bauer, and Reinhard Carius. Study of thin-film silicon solar cell back reflectors and potential of detached reflectors. *Energy Procedia*, 10:106–110, 2011. ISSN 1876-6102. doi: 10.1016/j.egypro.2011.10.161.

[17] Yi Zou, Xing Sheng, Kun Xia, Huayu Fu, and Juejun Hu. Parasitic loss suppression in photonic and plasmonic photovoltaic light trapping structures. *Opt. Express*, 22(S4):A1197–A1202, Jun 2014. doi: 10.1364/OE.22.0A1197.

[18] Reyu Sakakibara, Veronika Stelmakh, Walker R. Chan, Robert D. Geil, Stephan Krämer, Timothy Savas, Michael Ghebrebrhan, John D. Joannopoulos, Marin Soljačić, and Ivan Čelanović. A high-performance, metallodielectric 2d photonic crystal for thermophotovoltaics. *Solar Energy Materials and Solar Cells*, 238:111536, 2022. ISSN 0927-0248. doi: https://doi.org/10.1016/j.solmat.2021.111536.

[19] Hong-Yu Pan, Xin-Lin Xia, and Xue Chen. Multi-field coupled analysis of thermal and opto-electrical conversion in ingaas thermophotovoltaics. *Solar Energy Materials and Solar Cells*, 279:113242, 2025. ISSN 0927-0248. doi: https://doi.org/10.1016/j.solmat.2024.113242.

[20] Bibekananda Nath, Kawshik Nath, and Ahmed Zubair. Multilayer all-oxide polarization-independent narrowband emitter: A step towards the future thermophotovoltaic applications. In *TENCON 2025 - 2025 IEEE Region 10 Conference (TENCON)*, pages 1643–1647, 2025. doi: 10.1109/TENCON66050.2025.11375554.

[21] Yuchun Cao, Heng Zhang, Ning Chen, Haotuo Liu, Yongtao Feng, and Xiaohu Wu. A tungsten-based metamaterial emitter for solar thermophotovoltaic systems. *Physical Chemistry Chemical Physics*, 26(18):13909–13914, 05 2024. ISSN 1463-9076. doi: 10.1039/d4cp00210e.

[22] AA Khairul Azri, MS Mohd Jasni, SF Wan Muhamad Hatta, MA Islam, Y Abdul Wahab, S Mekhilef, and PJ Ker. Advancement in thermophotovoltaic technology and nanoparticle incorporation for power generation. *Solar Energy*, 259:279–297, 2023. doi: https://doi.org/10.1016/j.solener.2023.05.018.

[23] Kevin A Arpin, Mark D Losego, Andrew N Cloud, Hailong Ning, Justin Mallek, Nicholas P Sergeant, Linxiao Zhu, Zongfu Yu, Berç Kalanyan, Gregory N Parsons, et al. Three-dimensional self-assembled photonic crystals with high temperature stability for thermal emission modification. *Nature communications*, 4(1):2630, 2013.

[24] Ziyi Zhou, Xiao Peng, Weiyan Lü, Shouhua Yang, Haonan Li, Hongbo Guo, and Jianqiang Wang. Ultra-high temperature oxidation resistant refractory high entropy alloys fabricated by laser melting deposition: Al concentration regulation and oxidation mechanism. *Corrosion Science*, 224:111537, 2023. ISSN 0010-938X. doi: https://doi.org/10.1016/j.corsci.2023.111537.

[25] Jiawei Song, Zihao He, Chao Shen, Jie Zhu, Zhimin Qi, Xing Sun, Yizhi Zhang, Juncheng Liu, Xinghang Zhang, Xiulin Ruan, et al. Design of all-oxide multilayers with high-temperature stability toward future thermophotovoltaic applications. *Advanced Materials Interfaces*, 11(5):2300733, 2024.

[26] Junho Yoon, Ming Zhou, Md. Alamgir Badsha, Tae Young Kim, Young Chul Jun, and Chang Kwon Hwangbo. Broadband epsilon-near-zero perfect absorption in the near-infrared. *Scientific Reports*, 5(1):12788, Aug 2015. ISSN 2045-2322. doi: 10.1038/srep12788.

[27] Khant Minn, Aleksei Anopchenko, Jingyi Yang, and Ho Wai Howard Lee. Excitation of epsilon-near-zero resonance in ultra-thin indium tin oxide shell embedded nanostructured optical fiber. *Scientific Reports*, 8(1):2342, Feb 2018. ISSN 2045-2322. doi: 10.1038/s41598-018-19633-2.

[28] Thang Duy Dao, Anh Tung Doan, Dang Hai Ngo, Kai Chen, Satoshi Ishii, Akemi Tamanai, and Tadaaki Nagao. Selective thermal emitters with infrared plasmonic indium tin oxide working in the atmosphere. *Opt. Mater. Express*, 9(6):2534–2544, Jun 2019. doi: 10.1364/OME.9.002534.

[29] Otto J. Gregory, Qing Luo, and Everett E. Crisman. High temperature stability of indium tin oxide thin films. *Thin Solid Films*, 406(1):286–293, 2002. ISSN 0040-6090. doi: 10.1016/S0040-6090(01)01773-4.

[30] Jiwoong Kim, Sujan Shrestha, Maryam Souri, John G. Connell, Sungkyun Park, and Ambrose Seo. High-temperature optical properties of indium tin oxide thin-films. *Scientific Reports*, 10(1):12486, Jul 2020. ISSN 2045-2322. doi: 10.1038/s41598-020-69463-4.

[31] Haiou Li, Lei Guo, Xingpeng Liu, Tangyou Sun, Qi Li, Fabi Zhang, Gongli Xiao, Tao Fu, and Yonghe Chen. High temperature conductive stability of indium tin oxide films. *Frontiers in Materials*, Volume 7 - 2020, 2020. ISSN 2296-8016. doi: 10.3389/fmats.2020.00113.

[32] Esther López, Irene Artacho, and Alejandro Datas. Thermophotovoltaic conversion efficiency measurement at high view factors. *Solar Energy Materials and Solar Cells*, 250:112069, 2023. ISSN 0927-0248. doi: 10.1016/j.solmat.2022.112069.

[33] Alina LaPotin, Kevin L. Schulte, Myles A. Steiner, Kyle Buznitsky, Colin C. Kelsall, Daniel J. Friedman, Eric J. Tervo, Ryan M. France, Michelle R. Young, Andrew Rohskopf, Shomik Verma, Evelyn N. Wang, and Asegun Henry. Thermophotovoltaic efficiency of 40%. *Nature*, 604(7905):287–291, Apr 2022. ISSN 1476-4687. doi: 10.1038/s41586-022-04473-y.

[34] Ignacio Del Villar, Carlos R. Zamarre no, Miguel Hernaez, Francisco J. Arregui, and Ignacio R. Matias. Generation of lossy mode resonances with absorbing thin-films. *J. Lightwave Technol.*, 28(23): 3351–3357, Dec 2010. doi: 10.1109/JLT.2010.2082492.

[35] Stefano D'Elia, Nicola Scaramuzza, Federica Ciuchi, Carlo Versace, Giuseppe Strangi, and Roberto Bartolino. Ellipsometry investigation of the effects of annealing temperature on the optical properties of indium tin oxide thin films studied by drude–lorentz model. *Applied Surface Science*, 255(16):7203–7211, 2009. ISSN 0169-4332. doi: 10.1016/j.apsusc.2009.03.064.

[36] Juhn-Jong Lin and Zhi-Qing Li. Electronic conduction properties of indium tin oxide: Single-particle and many-body transport. *Journal of Physics: Condensed Matter*, 26(34):343201, 2014. ISSN 0953-8984. doi: 10.1088/0953-8984/26/34/343201.

[37] Gerardo Silva-Oelker, Juliana Jaramillo Fernández, and Nelson Toledo. Numerical study of high-temperature, disk-based tungsten and molybdenum thermophotovoltaic selective thermal emitters. *Optics Express*, 33(4):6953–6965, 2025. doi: 10.1364/OE.545130.

[38] Fangzhou Zhao, Ding Wang, Haotuo Liu, Xiaohu Wu, and Haoqiang Ai. Multilayer fabry-pérot selective emitter for efficient thermophotovoltaic energy conversion. *Case Studies in Thermal Engineering*, 84: 108320, 2026. ISSN 2214-157X. doi: 10.1016/j.csite.2026.108320.

[39] Heng Li, Jialu Tian, Shujian Sun, and Shiquan Shan. Design of efficient thermophotovoltaic system based on meta-material narrowband emitter for space power supply. *Thermal Science*, 28(1 Part A): 51–63, 2024. doi: 10.2298/TSCI221125087L.

[40] Meiya Rong, Kaixia Xu, Kezhang Shi, and Chengping Yin. Dual-polarization narrowband thermal vertical emitter with ultrahigh directionality. *Journal of Applied Physics*, 139(5), 2026. doi: 10.1063/5.0308301.

[41] Noriyuki Hasuike, Takeshi Maeda, and Minoru Takeda. Fabrication of ITO diffraction grating structure for infrared plasmonics by thermal nanoimprint lithography. *Optical Review*, 29:450–455, 2022. doi: 10.1007/s10043-022-00759-8.

[42] Vincent Vandalon and W. M. M. Erwin Kessels. Initial growth study of atomic-layer deposition of $Al_2O_3$ by vibrational sum-frequency generation. *Langmuir*, 35(32):10374–10382, 2019. doi: 10.1021/acs.langmuir.9b01600.

[43] Xue Yang, Dong-Pyo Kim, Doo-Seung Um, Gwan-Ha Kim, and Chang-Il Kim. Temperature dependence on dry etching of $Al_2O_3$ thin films in $BCl_3/Cl_2$/Ar plasma. *Journal of Vacuum Science & Technology A*, 27(4):821–825, 2009. doi: 10.1116/1.3086642.

[44] Ho-Won Yoon, Seung-Min Shin, Seong-Yong Kwon, Hyun-Min Cho, Sang-Gab Kim, and Mun-Pyo Hong. One-step etching characteristics of ITO/Ag/ITO multilayered electrode in high-density and high-electron-temperature plasma. *Materials*, 14(8):2025, 2021. doi: 10.3390/ma14082025.

# Supplementary Material for Surrogate-Assisted Inverse Design and Temperature-Dependent Electrothermal Analysis of All-Oxide Narrowband Emitter for Thermophotovoltaic Energy Conversion

Bibekananda Nath[a,b], Kawshik Nath[a,b], Ahmed Zubair[a]

[a] *Department of Electrical and Electronic Engineering, Bangladesh University of Engineering and Technology, Dhaka, Bangladesh*
[b] *Department of Electrical and Electronic Engineering, Chittagong University of Engineering and Technology, Chattogram, Bangladesh*

---

---

## S1. FDTD Design Library and Spectral Metrics

The optical design library was generated from FDTD-computed emittance spectra over 400–4000 nm. The geometric variables were the structural period $P$, ITO thickness $t_{\mathrm{ITO}}$, and $\mathrm{Al_2O_3}$ thickness $t_{\mathrm{Al_2O_3}}$. The discrete period values used in the library were

$$P = \{340, 350, 360, 370, 380, 400, 420, 440, 460\}\ \mathrm{nm}. \tag{S1}$$

The ITO thickness was sampled from 150 to 350 nm in 5 nm increments, and the $\mathrm{Al_2O_3}$ thickness was sampled from 5 to 350 nm in 5 nm increments. The resulting library contained 25,830 raw geometric combinations; after removal of invalid or nonphysical metric entries, 25,340 samples were retained for surrogate modeling.

For each spectrum, the peak emittance and peak wavelength were obtained as

$$E_{\mathrm{peak}} = \max_{\lambda \in [400,4000]} E(\lambda), \qquad \lambda_{\mathrm{peak}} = \arg \max_{\lambda \in [400,4000]} E(\lambda). \tag{S2}$$

The FWHM was calculated from the interpolated half-maximum crossings surrounding the dominant peak. Spectral concentration was evaluated over a peak-centered band

$$\mathcal{B} = [\lambda_{\mathrm{peak}} - 50, \lambda_{\mathrm{peak}} + 50]\ \mathrm{nm}, \tag{S3}$$

with in-band emission represented as follows:

$$f_{\mathrm{in}} = \frac{\int_{\mathcal{B}} E(\lambda)\, d\lambda}{\int_{400}^{4000} E(\lambda)\, d\lambda}, \tag{S4}$$

and out of band emission as follows:

$$R_{\mathrm{side}} = \frac{\displaystyle\int_{400}^{4000} E(\lambda)\, d\lambda - \int_{\mathcal{B}} E(\lambda)\, d\lambda}{\displaystyle\int_{\mathcal{B}} E(\lambda)\, d\lambda} = \frac{1 - f_{\mathrm{in}}}{f_{\mathrm{in}}}. \tag{S5}$$

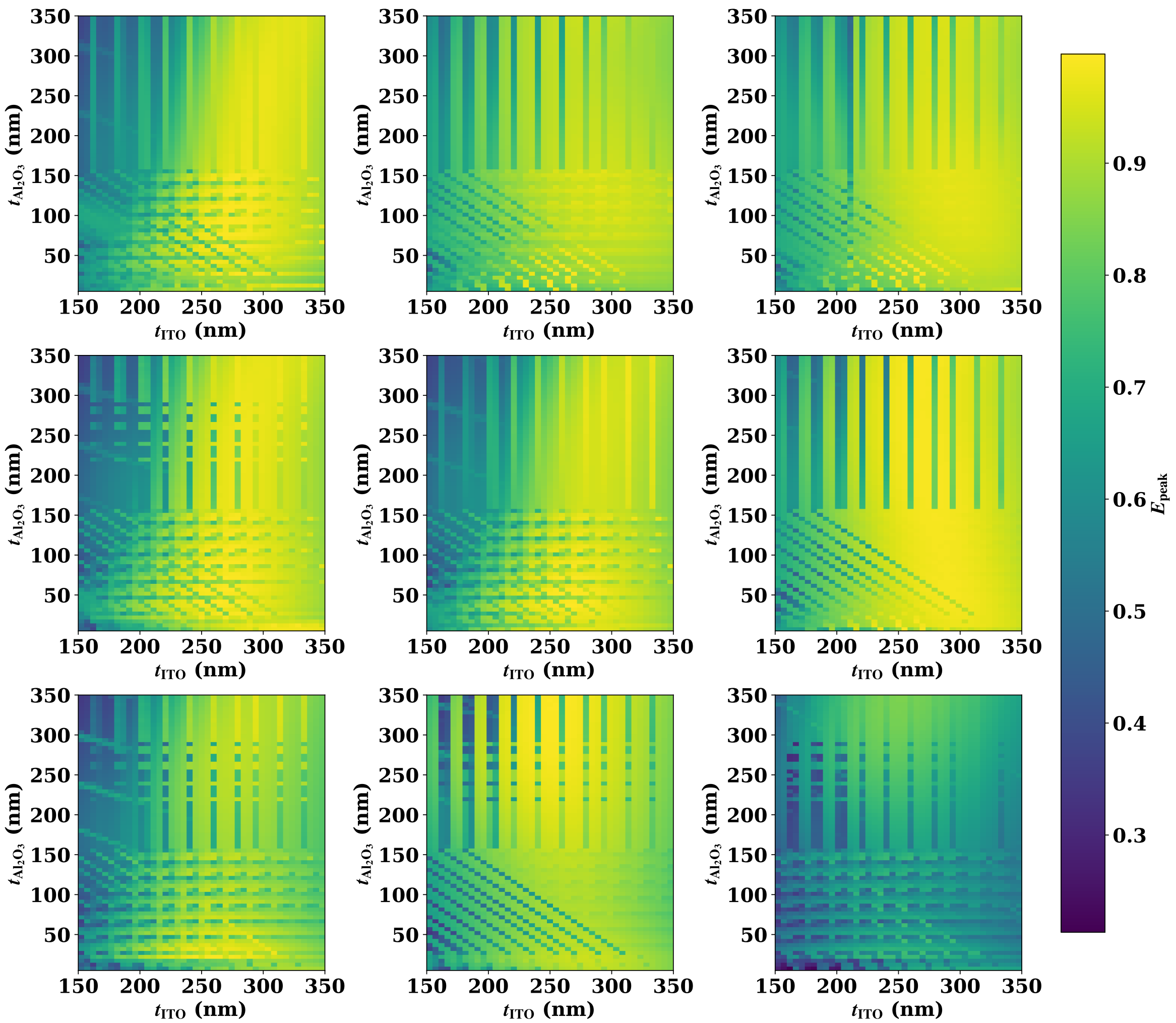

Figure S1: Period-resolved peak-emittance maps of the FDTD design library as functions of $t_{\mathrm{ITO}}$ and $t_{\mathrm{Al_2O_3}}$. Each panel corresponds to one discrete structural period.

Figure S1 shows that high-emittance solutions occupy structured and nonuniform regions of the geometric design space. The strong dependence on the combined values of $P$, $t_{\mathrm{ITO}}$, and $t_{\mathrm{Al_2O_3}}$ supports the use of a multivariable surrogate-assisted search rather than independent one-dimensional sweeps.

## S2. Surrogate-Assisted Bayesian Optimization

### *S2.1. Optimization surrogate and objective function*

Four ExtraTrees regressors were trained to predict $E_{\mathrm{peak}}$, $\lambda_{\mathrm{peak}}$, FWHM, and $f_{\mathrm{in}}$ from

$$\mathbf{x} = (P, t_{\mathrm{ITO}}, t_{\mathrm{Al_2O_3}}). \tag{S6}$$

Each optimization surrogate used 1200 trees, a minimum of two samples per leaf, a random state of zero, and parallel execution across all available CPU cores. FWHM was modeled as log(FWHM) and $R_{\mathrm{side}}$ was determined from predicted $f_{\mathrm{in}}$ according to eqation S5. The training weights were

$$w_i = 1 + 3\,\mathbb{I}(t_{\mathrm{Al_2O_3},i} \leq 20 \text{ nm}) + 2\,\mathbb{I}(E_{\mathrm{peak},i} \geq 0.95)\,, \tag{S7}$$

where $\mathbb{I}(\cdot)$ denotes the indicator function.

For feasible candidates, the scalar objective implemented in the optimization code was

$$J(\mathbf{x}) = 0.01\,\widehat{\mathrm{FWHM}}(\mathbf{x}) - 3.00\,\widehat{E}_{\mathrm{peak}}(\mathbf{x}) - 1.00\,\widehat{f}_{\mathrm{in}}(\mathbf{x}), \tag{S8}$$

subject to $\widehat{E}_{\mathrm{peak}} \geq 0.90$. Candidates below the peak-emittance threshold were assigned

$$J(\mathbf{x}) = 10^6 + 1000\left(0.90 - \widehat{E}_{\mathrm{peak}}(\mathbf{x})\right)^2, \qquad \widehat{E}_{\mathrm{peak}} < 0.90. \tag{S9}$$

The continuous search bounds were $340 \leq P \leq 460$ nm, $150 \leq t_{\mathrm{ITO}} \leq 350$ nm, and $5 \leq t_{\mathrm{Al_2O_3}} \leq 350$ nm. The search was initialized with $\mathbf{x}_0 = (400, 255, 5)$ nm, followed by 180 optimizer calls with 30 initial exploratory points and `random_state=0`. The saved optimization history contained 181 rows including the pre-evaluated seed, of which 143 satisfied $E_{\mathrm{peak}} \geq 0.90$.

*S2.2. Expected-Improvement search*

The Bayesian optimizer used a Gaussian-process (GP) model for the scalar objective generated from the ExtraTrees predictions. For an unexplored geometry,

$$J(\mathbf{x}) \sim \mathcal{N}\left[\mu_J(\mathbf{x}), \sigma_J^2(\mathbf{x})\right]. \tag{S10}$$

For minimization, Expected Improvement was evaluated as

$$EI(\mathbf{x}) = [J_{\mathrm{best}} - \mu_J(\mathbf{x})]\,\Phi(z) + \sigma_J(\mathbf{x})\phi(z), \tag{S11}$$

where

$$z = \frac{J_{\mathrm{best}} - \mu_J(\mathbf{x})}{\sigma_J(\mathbf{x})}, \tag{S12}$$

and $\Phi$ and $\phi$ are the standard-normal cumulative-distribution and probability-density functions, respectively. The next candidate was selected from

$$\mathbf{x}_{\mathrm{next}} = \arg\max_{\mathbf{x}} EI(\mathbf{x}). \tag{S13}$$

The selected geometry was evaluated by the ExtraTrees surrogates, converted to $J(\mathbf{x}_{\mathrm{next}})$, appended to the GP history, and the process was repeated sequentially.

Table S1: Surrogate-model and Bayesian-optimization settings used in the inverse-design workflow.

| **Setting** | **Value** | **Implementation detail** |
|---|---|---|
| Raw spectral-metric samples | 25,830 | Nine periods × 41 ITO thicknesses × 70 $Al_2O_3$ thicknesses |
| Valid samples after cleaning | 25,340 | Used for continuous-surrogate training |
| Discrete periods | 340, 350, 360, 370, 380, 400, 420, 440, 460 nm | FDTD design library |
| ITO-thickness range | 150–350 nm, 5 nm step | 41 values |
| $Al_2O_3$-thickness range | 5–350 nm, 5 nm step | 70 values |
| Spectral range | 400–4000 nm | Metric-extraction interval |
| In-band half-width | 50 nm | Peak-centered ±50 nm interval |
| Optimization surrogate | ExtraTreesRegressor | One model per spectral metric |
| Optimization-surrogate trees | 1200 | Minimum samples per leaf = 2 |
| FWHM transform | log(FWHM) | Inverted after prediction |
| Feasibility threshold | $E_{\mathrm{peak}} \geq 0.90$ | Applied before objective ranking |
| Objective weights | $w_{\mathrm{FWHM}} = 0.01$, $w_{\mathrm{peak}} = 3.00$, $w_{\mathrm{in}} = 1.00$ | Eq. (S8) |
| Seed geometry | $(400, 255, 5)$ nm | $(P, t_{\mathrm{ITO}}, t_{\mathrm{Al_2O_3}})$ |
| BO calls | 180 | `n_calls=180` |
| Initial exploratory points | 30 | `n_initial_points=30` |
| Random seed | 0 | `random_state=0` |
| Recorded BO rows | 181 | Includes the pre-evaluated seed |
| Feasible BO rows | 143 | $E_{\mathrm{peak}} \geq 0.90$ |
| Validation split | 80/20 | `test_size=0.2`, `random_state=0` |
| Validation-model trees | 600 | Minimum samples per leaf = 2 |
| Permutation repeats | 8 | Held-out test set |

Table S1 consolidates the numerical settings used for surrogate training, validation, and continuous Bayesian optimization. These settings correspond to the implementation used to generate the optimization and validation results reported in the main manuscript.

## S3. Optimization Convergence and Surrogate Validation

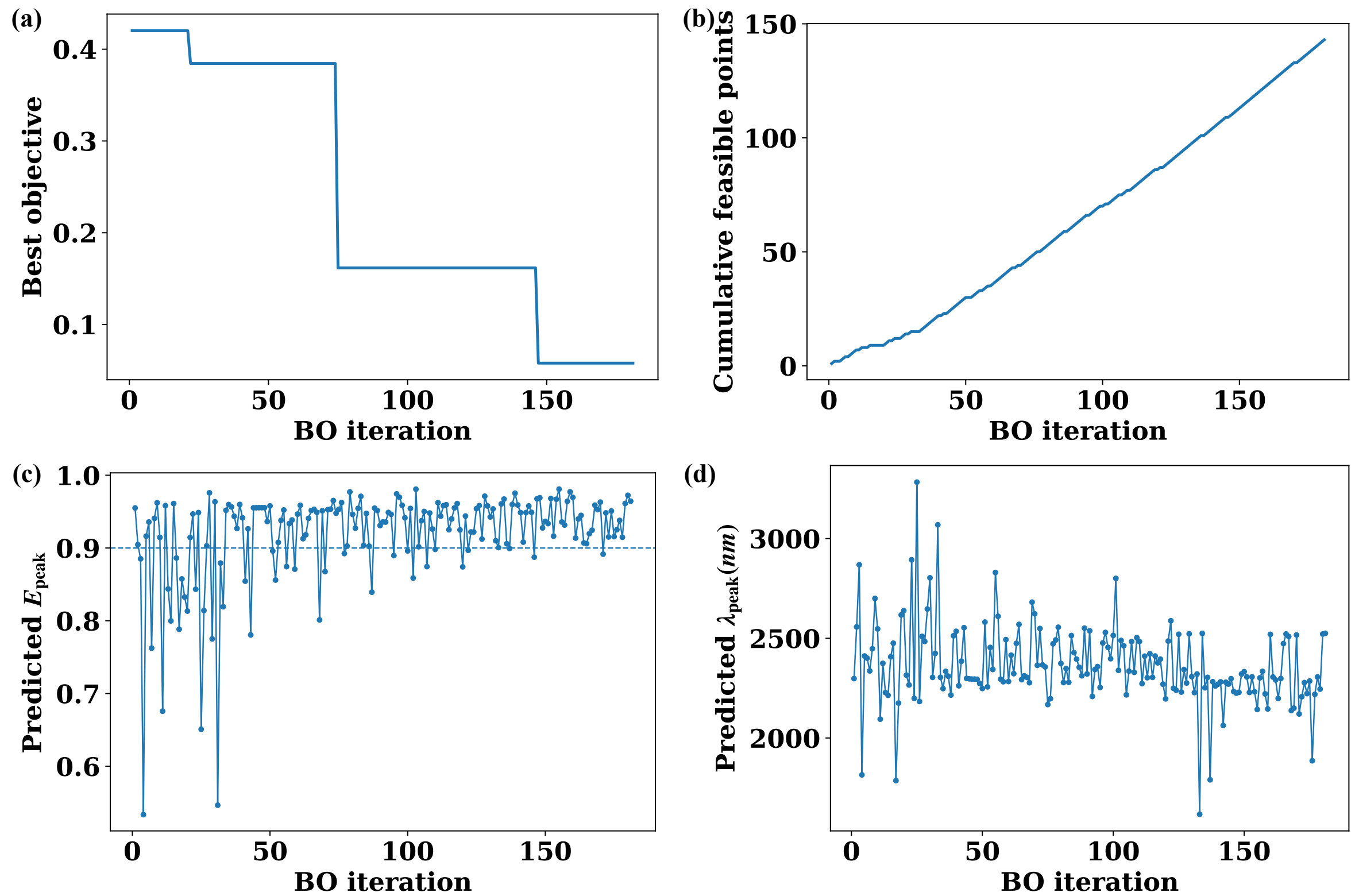

Figure S2: Bayesian-optimization evolution: (a) running minimum of the penalty-augmented objective; (b) cumulative number of feasible candidates; (c) surrogate-predicted $E_{\text{peak}}$ with the 0.90 feasibility threshold; and (d) predicted $\lambda_{\text{peak}}$ versus iteration.

Figure S2 shows a progressive reduction of the best objective together with continued discovery of feasible candidates. The variation in predicted peak wavelength indicates that the search continued to sample multiple spectral solutions rather than collapsing immediately onto a single region.

The held-out ExtraTrees models were evaluated using MAE, RMSE, and $R^2$. The resulting validation scores are summarized in Table S2.

Table S2: Predictive performance of the ExtraTrees validation models on the held-out test set.

| Target metric | MAE | RMSE | $R^2$ |
|---|---|---|---|
| $E_{\text{peak}}$ | 0.0236 | 0.0439 | 0.918 |
| FWHM | 15.6 nm | 32.6 nm | 0.837 |
| $\lambda_{\text{peak}}$ | 97.4 nm | 218 nm | 0.775 |
| $f_{\text{in}}$ | 0.00507 | 0.011 | 0.664 |

Table S2 shows the strongest predictive agreement for $E_{\text{peak}}$, followed by FWHM. The lower $R^2$ values for $\lambda_{\mathrm{peak}}$ and $f_{\mathrm{in}}$ are consistent with the greater sensitivity of peak selection and integrated spectral concentration to changes in the detailed spectral shape.

## S4. FDTD Numerical Settings

The principal numerical settings used in the FDTD calculations are listed in Table S3.

Table S3: Numerical settings used in the FDTD simulation of the $Al_2O_3$/ITO/sapphire emitter.

| **Simulation quantity** | **Value** | **Remarks** |
|---|---|---|
| Software | Ansys Lumerical FDTD | Full-wave optical simulation |
| Wavelength range | 400–4000 nm | Matches the metric-extraction interval |
| Source | Broadband plane wave | Propagation along $-z$ |
| Polarization | TE or TM, according to simulation case | Normal and angular-response calculations |
| $x$ boundary condition | Periodic | Lateral boundary |
| $y$ boundary condition | Periodic | Lateral boundary |
| $z$ boundary condition | Perfectly matched layer (PML) | Normal boundary |
| Global mesh accuracy | Nonuniform mesh setting 8 | High-accuracy setting |
| Local mesh, $x$ | 5 nm | Uniform mesh |
| Local mesh, $y$ | 5 nm | Uniform mesh |
| Local mesh, $z$ | 2.5 nm | Uniform mesh |
| Simulation time | 1000 fs | – |
| Auto-shutoff minimum | $10^{-7}$ | Convergence criterion |
| Spectral points | 500 | Wavelength sampling |
| Reflection monitor | Above the structure | Reflection extraction |
| Transmission monitor | Below the structure | Transmission extraction |
| Material fit tolerance | 0.1 | Maximum coefficients = 6 |

Table S3 records the source, boundary conditions, spatial discretization, spectral sampling, and monitor settings used to generate the optical design library and to verify the selected emitter response.

## S5. Temperature-Dependent Optical Model of ITO

The temperature-dependent ITO model combined the room-temperature optical constants of Del Villar *et al.* [1] with a temperature-dependent modification of the Drude damping. At 300 K, the reference complex permittivity was calculated from

$$\tilde{\varepsilon}_{\mathrm{DV}}(\lambda, 300) = \left[n_{300}(\lambda) + ik_{300}(\lambda)\right]^2 . \tag{S14}$$

*S5.1. Room-temperature Drude–Lorentz fit*

The reference response was fitted in the photon-energy domain using

$$\tilde{\varepsilon}_{\mathrm{fit}}(E) = \varepsilon_\infty - \frac{A_D}{E^2 + iB_D E} + \frac{A_L}{E_L^2 - E^2 - iB_L E}, \qquad E = \frac{hc}{\lambda}. \tag{S15}$$

The model parameters were obtained by simultaneous fitting of $n$ and $k$ using

$$\chi^2 = \sum_{j=1}^{N_\lambda} \left[ \frac{n_{\mathrm{DL}}(\lambda_j) - n_{\mathrm{DV}}(\lambda_j)}{s_n} \right]^2 + \sum_{j=1}^{N_\lambda} \left[ \frac{k_{\mathrm{DL}}(\lambda_j) - k_{\mathrm{DV}}(\lambda_j)}{s_k} \right]^2, \tag{S16}$$

where $s_n$ and $s_k$ balance the two fitting contributions. The fitted response gave $\mathrm{RMSE}_n =$ 0.00206 and $\mathrm{RMSE}_k = 0.00534$. The fitted parameters are listed in Table S4.

Table S4: Room-temperature Drude–Lorentz parameters and temperature-model constants for ITO.

| **Parameter** | **Value** | **Unit** |
|---|---|---|
| $\varepsilon_\infty$ | 2.9249205 | – |
| $A_D$ | 3.4867682 | $\mathrm{eV}^2$ |
| $B_D(300)$ | 0.024621825 | eV |
| $A_L$ | 24.953058 | $\mathrm{eV}^2$ |
| $B_L$ | 0.0032279307 | eV |
| $E_L$ | 5.0323017 | eV |
| $\Theta_D$ | 1000 | K |
| $f_{\mathrm{res}}$ | 0.909 | – |
| $\mathrm{RMSE}_n$ | 0.00206 | – |
| $\mathrm{RMSE}_k$ | 0.00534 | – |

Table S4 shows the fitted room-temperature dispersion parameters together with the two constants used to scale the Drude damping with temperature.

*S5.2. Temperature-dependent Drude damping*

The fitted damping was separated conceptually into residual and electron–phonon contributions,

$$B_D(T) = B_{D,\mathrm{res}} + B_{D,\mathrm{e-ph}}(T). \tag{S17}$$

The electron–phonon contribution was represented by the normalized Bloch–Grüneisen function

$$\Phi_{\mathrm{BG}}(T, \Theta_D) = \left( \frac{T}{\Theta_D} \right)^5 \int_0^{\Theta_D/T} \frac{x^5 e^x}{(e^x - 1)^2}\, dx, \tag{S18}$$

leading to

$$B_D(T) = B_D(300)\left[f_{\text{res}} + (1-f_{\text{res}})\frac{\Phi_{\text{BG}}(T,\Theta_D)}{\Phi_{\text{BG}}(300,\Theta_D)}\right]. \tag{S19}$$

An effective transport Debye temperature of $\Theta_D = 1000$ K was adopted from the transport analysis of Lin and Li [3]. Using the approximate room-temperature-to-low-temperature resistivity ratio $\rho(300)/\rho(25) \approx 1.1$ and treating $\rho(25)$ as the residual contribution gives

$$\frac{B_{D,\text{e-ph}}(300)}{B_{D,\text{res}}} \approx 0.1, \qquad f_{\text{res}} = \frac{1}{1+0.1} \approx 0.909. \tag{S20}$$

Thus, $f_{\text{res}}$ is a literature-informed estimate of the residual fraction rather than a universal ITO material constant.

*S5.3. Temperature-dependent permittivity and optical constants*

Only the Drude damping parameter was varied with temperature. The temperature-modified Drude contribution was

$$\tilde{\varepsilon}_D(E,T) = -\frac{A_D(300)}{E^2 + iB_D(T)E}. \tag{S21}$$

The complete permittivity was constructed as

$$\tilde{\varepsilon}(E,T) = \tilde{\varepsilon}_{\text{DV}}(E,300) + \tilde{\varepsilon}_D(E,T) - \tilde{\varepsilon}_D(E,300), \tag{S22}$$

which exactly recovers the imported Del Villar response at 300 K. Writing $\tilde{\varepsilon} = \varepsilon_1 + i\varepsilon_2$, the corresponding optical constants were calculated as

$$n(\lambda,T) = \sqrt{\frac{\sqrt{\varepsilon_1^2+\varepsilon_2^2}+\varepsilon_1}{2}}, \tag{S23}$$

$$k(\lambda,T) = \sqrt{\frac{\sqrt{\varepsilon_1^2+\varepsilon_2^2}-\varepsilon_1}{2}}. \tag{S24}$$

The elevated-temperature optical constants are therefore model-predicted responses based on temperature-dependent free-carrier damping. Carrier-density changes, effective-mass changes, interband shifts, crystallinity changes, and oxygen-vacancy redistribution were not independently parameterized because matching temperature-resolved data for the same ITO film were unavailable. The agreement at 300 K reflects preservation of the reference response by construction and is not an independent validation of the elevated-temperature model.

## S6. InGaAs Material and Electrothermal Parameters

The optimized emitter was coupled to an $In_{0.73}Ga_{0.27}As$ TPV cell. The composition-dependent semiconductor parameters adopted from Lee *et al.* [4] are summarized in Table S5.

Table S5: Reference parameters for $In_{0.73}Ga_{0.27}As$ at 300 K from Lee *et al.* [4].

| **Parameter** | **Symbol** | **Value** | **Unit** |
|---|---|---|---|
| Bandgap energy | $E_g$ | 0.564 | eV |
| Electron effective density of states | $N_c$ | $1.453 \times 10^{17}$ | $cm^{-3}$ |
| Hole effective density of states | $N_v$ | $7.226 \times 10^{18}$ | $cm^{-3}$ |
| Static relative permittivity | $\varepsilon_r$ | 14.37 | – |
| Intrinsic carrier concentration | $n_i$ | $2.03 \times 10^{13}$ | $cm^{-3}$ |
| Electron mobility | $\mu_e$ | $1.09 \times 10^3$ | $cm^2V^{-1}s^{-1}$ |
| Hole mobility | $\mu_h$ | $1.48 \times 10^2$ | $cm^2V^{-1}s^{-1}$ |
| Electron diffusion coefficient | $D_e$ | 28.1 | $cm^2s^{-1}$ |
| Hole diffusion coefficient | $D_h$ | 3.81 | $cm^2s^{-1}$ |
| Electron lifetime | $\tau_e$ | $1.11 \times 10^{-11}$ | s |
| Hole lifetime | $\tau_h$ | $7.37 \times 10^{-8}$ | s |
| Electron diffusion length | $L_e$ | $1.76 \times 10^{-7}$ | m |
| Hole diffusion length | $L_h$ | $5.30 \times 10^{-6}$ | m |

Table S5 provides the reference material quantities used to define the InGaAs semiconductor response. The device-specific geometry, doping, recombination, and thermal parameters used in the coupled cell calculation are listed separately in Table S6.

Table S6: TPV-cell and electrothermal parameters used in the coupled simulations.

| Parameter | Symbol/region | Value | Unit/condition |
|---|---|---|---|
| $n^{++}$ layer thickness | $t_{n^{++}}$ | 0.4 | $\mu$m |
| $n^{++}$ doping concentration | $N_{D,n^{++}}$ | $9 \times 10^{19}$ | $cm^{-3}$ |
| $p$ layer thickness | $t_p$ | 6 | $\mu$m |
| $p$ doping concentration | $N_{A,p}$ | $1.36 \times 10^{16}$ | $cm^{-3}$ |
| $p^{++}$ layer thickness | $t_{p^{++}}$ | 0.2 | $\mu$m |
| $p^{++}$ doping concentration | $N_{A,p^{++}}$ | $1 \times 10^{20}$ | $cm^{-3}$ |
| Ag front-contact thickness | $t_{\text{Ag}}$ | 0.5 | $\mu$m |
| Al back-contact thickness | $t_{\text{Al}}$ | 1 | $\mu$m |
| Cu heat-sink thickness | $t_{\text{Cu}}$ | 42 | $\mu$m |
| Electron SRH lifetime | $\tau_n$ | $1.11 \times 10^{-11}$ | s |
| Hole SRH lifetime | $\tau_p$ | $7.37 \times 10^{-8}$ | s |
| Surface recombination velocity | – | $1 \times 10^{7}$ | $cm\,s^{-1}$ |
| Radiative recombination coefficient | $B$ | $2 \times 10^{-11}$ | $cm^3s^{-1}$ |
| Thermal conductivity of InGaAs | $k_{\text{InGaAs}}$ | 50 | $W\,m^{-1}K^{-1}$ |
| Thermal conductivity of Cu | $k_{\text{Cu}}$ | 316 | $W\,m^{-1}K^{-1}$ |
| Ambient/reference temperature | $T_{\text{amb}}$ | 300 | K |
| View factor | $F_{\text{view}}$ | 1 | – |
| Heat-transfer coefficient | $h$ | 75 | $W\,m^{-2}K^{-1}$ |
| Electrical mesh | – | 20–300 nm | Adaptive |
| Thermal mesh | – | 20–300 nm | Adaptive |
| Voltage sweep | – | 0–0.6 | V |
| Solver convergence | – | $10^{-4}$ absolute; $10^{-6}$ relative | Update criterion |

Table S6 summarizes the structural, electrical, recombination, thermal, and numerical settings used in the TPV-cell calculation. Heat removal was represented through convection from the Cu heat sink to the 300 K ambient using the listed heat-transfer coefficient.

## S7. Optimized structure validation and Mesh-Convergence Analysis

The optimized-emitter spectrum was validated using three mesh configurations to assess the sensitivity of the principal spectral response.

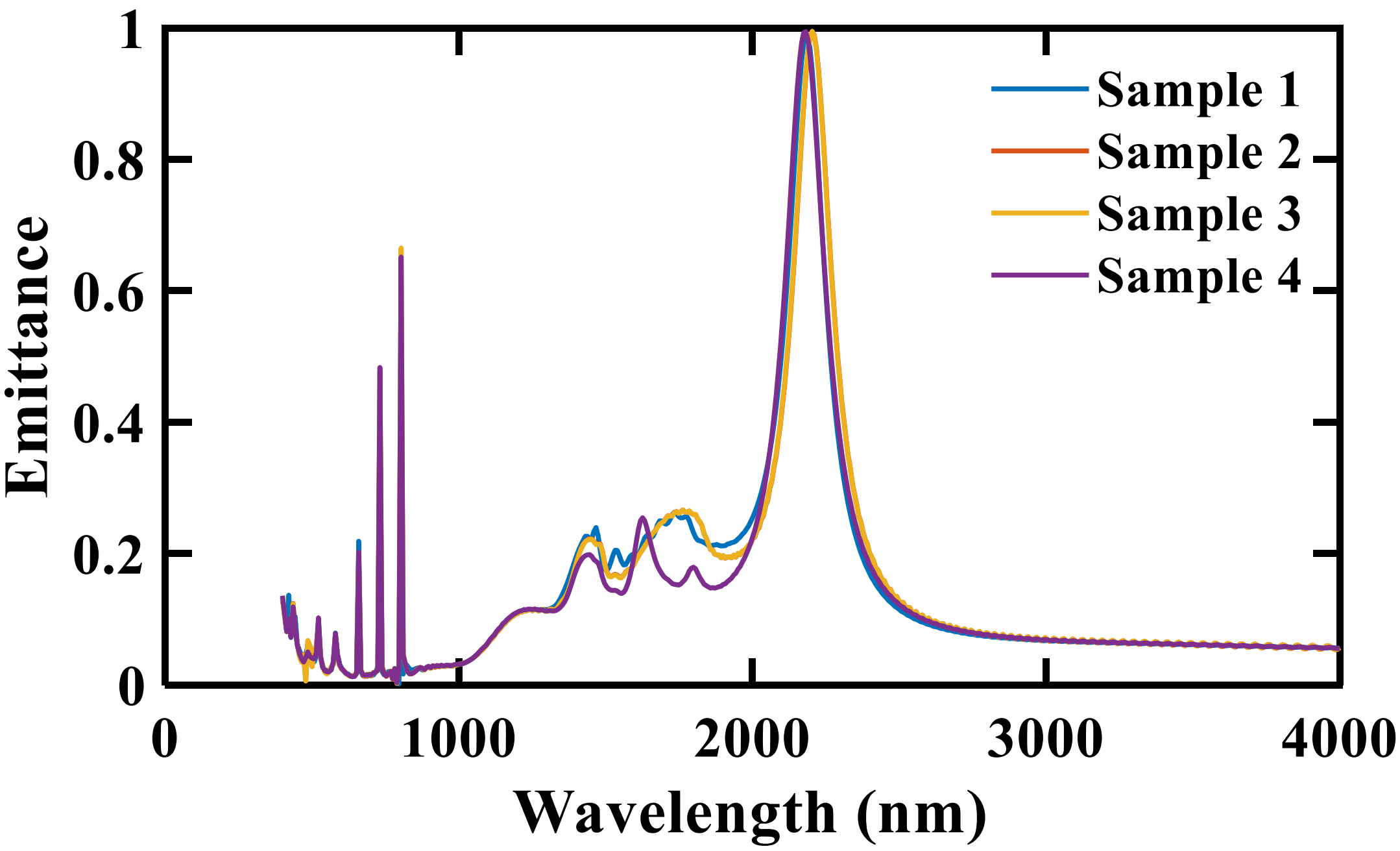

Figure S3: Spectral emittance of the optimized emitter obtained from surrogate-assisted optimized design and using three FDTD mesh configurations.

Figure S3 shows that the dominant narrowband resonance remains near the same wavelength with near-unity peak emittance across the investigated mesh configurations, while somewhat larger differences occur in linewidth and integrated spectral concentration.

Table S7: Optical performance obtained under the four mesh configurations.

| **Mesh** | **Mesh configuration** | $E_{\text{peak}}$ | $\lambda_{\text{peak}}$ (nm) | **FWHM (nm)** | $f_{\text{in}}$ |
|---|---|---|---|---|---|
| Sample 1 | Surrogate-assisted Inverse design Model | 0.9940 | 2181.96 | 169.29 | 0.172 |
| Sample 2 | Nonuniform mesh 8; uniform mesh $x = y = 5$ nm and $z = 2.5$ nm | 0.9931 | 2182.30 | 173.13 | 0.1715 |
| Sample 3 | Nonuniform mesh 2; uniform mesh $x = y = 5$ nm and $z = 2.5$ nm | 0.9890 | 2203.60 | 176.73 | 0.1813 |
| Sample 4 | Nonuniform mesh 8 | 0.9901 | 2183.4 | 173.97 | 0.1716 |

Table S7 lists the extracted peak and linewidth metrics for each mesh case. Relative changes were calculated with Sample 1 as the reference using

$$\Delta X(\%) = \frac{X_i - X_{\text{ref}}}{X_{\text{ref}}} \times 100. \tag{S25}$$

Table S8: Relative changes in the optical metrics with respect to Sample 1.

| **Mesh** | $\Delta E_{\text{peak}}$ (%) | $\Delta\lambda_{\text{peak}}$ (%) | $\Delta$FWHM (%) | $\Delta f_{\text{in}}$ (%) |
|---|---|---|---|---|
| Sample 1 | Reference | Reference | Reference | Reference |
| Sample 2 | −0.091 | +0.016 | +2.268 | +0.117 |
| Sample 3 | −0.503 | +0.992 | +4.4 | +5.838 |
| Sample 4 | −0.392 | +0.065 | +2.76 | +0.175 |

From Table S8, it can be observed that the optimized emitter response obtained from the inverse design method exhibits characteristics similar to those of the emitter structure analyzed via the FDTD method under different mesh conditions. More importantly, the reference values mostly match the emitter spectrum attributes obtained from the Sample 1 mesh setting used to prepare the dataset. Besides, it can be observed that various mesh settings provide almost identical results, with nonuniform mesh setting 2 providing a slightly larger deviation of only around 4.4% of FWHM and 5.8% in $f_{\text{in}}$ compared to the more refined mesh setting of Sample 1.

## References

[1] I. Del Villar, C. R. Zamarreño, M. Hernaez, F. J. Arregui, and I. R. Matias, "Generation of lossy mode resonances with absorbing thin-films," *Journal of Lightwave Technology*, vol. 28, no. 23, pp. 3351–3357, 2010.

[2] S. D'Elia, N. Scaramuzza, F. Ciuchi, C. Versace, G. Strangi, and R. Bartolino, "Ellipsometry investigation of the effects of annealing temperature on the optical properties of indium tin oxide thin films studied by Drude–Lorentz model," *Applied Surface Science*, vol. 255, no. 16, pp. 7203–7211, 2009.

[3] J.-J. Lin and Z.-Q. Li, "Electronic conduction properties of indium tin oxide: Single-particle and many-body transport," *Journal of Physics: Condensed Matter*, vol. 26, no. 34, p. 343201, 2014.

[4] H. J. Lee, M. M. A. Gamel, P. J. Ker, M. Z. Jamaludin, Y. H. Wong, K. S. Yap, J. R. Willmott, M. J. Hobbs, J. P. R. David, and C. H. Tan, "Deriving the absorption coefficients of lattice mismatched InGaAs using genetic algorithm," *Materials Science in Semiconductor Processing*, vol. 153, p. 107135, 2023.